\documentclass[aps,prl,preprint,nopacs,superscriptaddress]{revtex4}
\usepackage{amsmath}
\usepackage{amssymb}
\usepackage{graphicx}
\usepackage{hyperref}
\usepackage[utf8]{inputenc}
\newcommand{\beq}{\begin{equation*}}
\newcommand{\eeq}{\end{equation*}}
\newcommand{\s}{Co$_2$MnGa}

\begin{document}

\title{A three-dimensional magnetic topological phase}


\author{Ilya Belopolski} \email{ilyab@princeton.edu}
\affiliation{Laboratory for Topological Quantum Matter and Spectroscopy (B7), Department of Physics, Princeton University, Princeton, New Jersey 08544, USA}

\author{Daniel S. Sanchez}
\affiliation{Laboratory for Topological Quantum Matter and Spectroscopy (B7), Department of Physics, Princeton University, Princeton, New Jersey 08544, USA}

\author{Guoqing Chang}
\affiliation{Centre for Advanced 2D Materials and Graphene Research Centre, National University of Singapore, 6 Science Drive 2, 117546, Singapore}
\affiliation{Department of Physics, National University of Singapore, 2 Science Drive 3, 117546, Singapore}

\author{Kaustuv Manna}
\affiliation{Max Planck Institute for Chemical Physics of Solids, N\"othnitzer Stra{\ss}e 40, 01187 Dresden, Germany}


\author{Benedikt Ernst}
\affiliation{Max Planck Institute for Chemical Physics of Solids, N\"othnitzer Stra{\ss}e 40, 01187 Dresden, Germany}

\author{Su-Yang Xu}
\affiliation{Laboratory for Topological Quantum Matter and Spectroscopy (B7), Department of Physics, Princeton University, Princeton, New Jersey 08544, USA}

\author{Songtian S. Zhang}
\affiliation{Laboratory for Topological Quantum Matter and Spectroscopy (B7), Department of Physics, Princeton University, Princeton, New Jersey 08544, USA}

\author{Hao Zheng}
\affiliation{Laboratory for Topological Quantum Matter and Spectroscopy (B7), Department of Physics, Princeton University, Princeton, New Jersey 08544, USA}

\author{Jiaxin Yin}
\affiliation{Laboratory for Topological Quantum Matter and Spectroscopy (B7), Department of Physics, Princeton University, Princeton, New Jersey 08544, USA}

\author{Bahadur Singh}
\affiliation{Centre for Advanced 2D Materials and Graphene Research Centre, National University of Singapore, 6 Science Drive 2, 117546, Singapore}
\affiliation{Department of Physics, National University of Singapore, 2 Science Drive 3, 117546, Singapore}

\author{Guang Bian}
\affiliation{Department of Physics \& Astronomy, University of Missouri, Columbia, Missouri 65211, USA}

\author{Daniel Multer}
\affiliation{Laboratory for Topological Quantum Matter and Spectroscopy (B7), Department of Physics, Princeton University, Princeton, New Jersey 08544, USA}

\author{Xiaoting Zhou}
\affiliation{Centre for Advanced 2D Materials and Graphene Research Centre, National University of Singapore, 6 Science Drive 2, 117546, Singapore}
\affiliation{Department of Physics, National University of Singapore, 2 Science Drive 3, 117546, Singapore}

\author{Shin-Ming Huang}
\affiliation{Department of Physics, National Sun Yat-Sen University, Kaohsiung 804, Taiwan}

\author{Baokai Wang}
\affiliation{Department of Physics, Northeastern University, Boston, Massachusetts 02115, USA}

\author{Arun Bansil}
\affiliation{Department of Physics, Northeastern University, Boston, Massachusetts 02115, USA}



\author{Hsin Lin}
\affiliation{Centre for Advanced 2D Materials and Graphene Research Centre, National University of Singapore, 6 Science Drive 2, 117546, Singapore} \affiliation{Department of Physics, National University of Singapore, 2 Science Drive 3, 117546, Singapore}

\author{Claudia Felser}
\affiliation{Max Planck Institute for Chemical Physics of Solids, N\"othnitzer Stra{\ss}e 40, 01187 Dresden, Germany}

\author{M. Zahid Hasan} \email{mzhasan@princeton.edu}
\affiliation{Laboratory for Topological Quantum Matter and Spectroscopy (B7), Department of Physics, Princeton University, Princeton, New Jersey 08544, USA}
\affiliation{Princeton Institute for Science and Technology of Materials, Princeton University, Princeton, New Jersey, 08544, USA}
\affiliation{Lawrence Berkeley National Laboratory, Berkeley, CA 94720, USA}




\begin{abstract}
Topological materials are of great recent interest in physics. Although much research has focused on understanding topological insulators, topological superconductors, Weyl semimetals and the phenomena they exhibit, it remains an open question whether a three-dimensional magnet can be topological with exotic surface states. Here using density functional theory and angle-resolved photoemission spectroscopy we demonstrate a topological phase in a bulk magnet, Co$_2$MnGa. We find that in this compound topological line nodes cross the Fermi level and we further observe drumhead surface states stretching across the line nodes in the surface Brillouin zone. Our bulk and surface measurements suggest the experimental realization of the first topological magnet in three dimensions, as well as the first topological magnetic metal in any dimension. Our observation of topological line nodes explains the puzzlingly large anomalous Hall current observed in Co$_2$MnGa and suggests a method for generating highly spin-polarized currents in materials which are not half-metals. Taken together, our results suggest a rich, unconventional interplay between magnetism and topology.
\end{abstract}

\date{\today}
\maketitle



A crystal in a topological phase has an electronic band structure characterized by a topological invariant \cite{news, ZahidColloq, myARCMP}. The study of topological phases of matter over the course of the past decade has led to the discovery of many crystals with different topological invariants and fascinating electronic properties. Chief examples among these include the $\mathbb{Z}_2$ topological insulator \cite{ZahidColloq, David, QSH_Molenkamp_2007, Xia} and the Weyl semimetal \cite{myARCMP, TaAsUs, TaAsThem, TaAsChen, TaPUs, NbPme}. It is natural to ask if a magnet can host a topological phase. Theoretically, it is well-known that there exists a rich classification of time-reversal symmetry breaking topological phases \cite{schnyder}. A topological magnet may also exhibit new properties, such as topological invariants which can be flipped by changing the direction of magnetization of the sample. There are a few known examples of magnetic topological phases, notably the original integer quantum Hall effect \cite{QH} and the quantum anomalous Hall effect \cite{QAH}. However, these phases both live in two dimensions. In three dimensions, there is no clear experimental example of a topological magnetic ground state.


To search for a topological magnet it is reasonable to consider ferromagnets with high Curie temperature. Compared to other kinds of ordering, we expect a ferromagnetic order to be easiest to capture in \textit{ab initio} calculation. If the Curie temperature is large, we also expect a large spin-splitting in the band structure, which may be easier to measure in experiment. Given these considerations, it is natural to study the cobalt-based full Heusler compounds, Co$_2Y\hspace{-0.5mm}Z$ ($Y$ a transition metal, $Z$ a $p$-block element). These are typically ferromagnetic, with Curie temperature as high as $\sim 1120$ K in the case of Co$_2$FeSi \cite{FelserReview}. In addition, a large anomalous Hall effect has been observed in Co$_2$MnAl \cite{CML-AHE}, raising the possibility that topological objects in the band structure give rise to a large Berry curvature field. Indeed, Co$_2$TiSn and several other compounds in this family have recently been proposed as candidates for a magnetic topological phase \cite{GuoqingHeusler,ZhijunHeusler}. Moreover, a number of interesting effects relevant for spintronics have previously been observed in this family. Specifically, a magnetic tunnel junction (MTJ) built from Co$_2$MnSi showed a tunnel magnetoresistance (TMR) of up to 1995\% at 4.2 K \cite{MTJ-TMR}. Also, a current-perpendicular-to-plane giant magnetoresistance (CPP-GMR) of 175\% at 80 K was achieved using Co$_2$Fe$_{0.4}$Mn$_{0.6}$Si \cite{CPP-GMR}. MTJs built from Co$_2$FeSi and Co$_2$FeAl have been used as a platform for the tunnel magneto-Seebeck effect (TMS) \cite{TMS}, while Co$_2$TiAl is a promising candidate for the spin Seebeck effect \cite{SSE}. A topological invariant in combination with the magnetic properties or spintronics applications may give rise to new phenomena. However, the ground state topological properties of this class of materials are unknown from experiment.




Here we use density functional theory (DFT) and angle-resolved photoemission spectroscopy (ARPES) to demonstrate a magnetic topological phase in Co$_2$MnGa. Specifically, we observe that a pair of bulk bands crosses along curves in the bulk Brillouin zone. Comparing our results to calculation, we argue that these extended degeneracies are topological line nodes associated with a $\pi$ Berry phase. The line nodes are protected by mirror symmetry and require ferromagnetic order in the sense that broken time-reversal symmetry is necessary to achieve spin-polarized bands. We also observe a drumhead surface state stretching across the line nodes, consistent between ARPES and DFT. Our results suggest that we have observed the first three-dimensional topological magnetic phase and the first topological magnetic metal in any dimension. Based on our results, we propose a new approach to generating spin-polarized currents which doesn't rely on half-metallicity. More broadly, our work opens the study of a rich classification of three-dimensional magnetic topological phases and holds promise for observing new phenomena arising from the interplay of topological invariants and magnetic order.





Co$_2$MnGa takes the full Heusler crystal structure, with a cubic face-centered Bravais lattice, space group $Fm\bar{3}m$ (No. 225), Fig. \ref{Fig1}A. As noted above, it is ferromagnetic with Curie temperature $T_\textrm{C} \sim 687$ K, see the Supplementary Materials and Ref. \cite{CMGCurie}. In calculation, the electronic structure of Co$_2$MnGa shows a simple band crossing between one valence and one conduction band, both majority spin polarized, see the $K-\Gamma$ segment at binding energy $E_{\textrm{B}} \sim 0.1$ eV in Fig. \ref{Fig1}B. The minority spin shows a large irrelevant pocket at $\Gamma$, but is otherwise far in energy from the Fermi level $E_\textrm{F}$. This minority spin pocket is unimportant for what follows, but see a more complete discussion in the Supplementary Materials, Fig. S16 and the accompanying text.

We consider whether the spin majority band crossing can give rise to a topological phase. We note that space group $Fm\bar{3}m$ is highly symmetric, including a large set of mirror symmetries, $M_x: x \rightarrow -x$, $M_{xy}: (x,y) \rightarrow (y,x)$ and the analogous ones for other axes. In the Brillouin zone, a mirror symmetry corresponds to mirror planes on which all Bloch states can be classified by a mirror eigenvalue, block diagonalizing the Bloch Hamiltonian into subsystems of opposite mirror eigenvalues. Since these two subsystems are decoupled by the mirror symmetry, if the bands of the subsystems overlap, the band crossings are protected. If we imagine two decoupled band structures in a two-dimensional Brillouin zone, we can see that they will generically overlap on closed curves, forming line nodes, as illustrated by the yellow curve in Fig. \ref{Fig1}C. Away from the line nodes the bands disperse linearly, forming cones in the band structure. In this sense, line nodes are related to an ordinary Dirac or Weyl semimetal, but with the point nodes upgraded to line nodes by the addition of a mirror symmetry. Such line nodes are inherently topological objects, associated with a $\pi$ Berry phase on a contour integral along any closed momentum space path which loops around the line node, see Fig. \ref{Fig1}D and Ref. \cite{schnyder2}. We note that \textit{non-magnetic} line nodes have received some attention in the literature in materials without inversion symmetry or where spin-orbit coupling is negligible \cite{MadhabZrSiS, SchoopZrSiS, ChenHfSiS, GuangPbTaSe2}. We calculate the spin majority band crossing of \s\ and we find three families of line nodes, marked red, blue and yellow, copied within each family by the symmetries of the lattice, Fig. \ref{Fig1}E. All line nodes are twofold degenerate, meaning they arise from crossings of two singly-degenerate bands. These line nodes are confined in the mirror planes, but they disperse in energy, see the Supplementary Materials, Fig. S5. For a related discussion on the topological phase of Co$_2$MnGa in calculation, see our recent theory work \cite{CMG_thy}. Based on our theoretical analysis, we search for a magnetic topological line node in \s\ using ARPES.





We make some comments on our ARPES measurements on \s. We compare a measured Fermi surface, Fig. \ref{Fig2}A, with the \textit{ab initio} constant energy surface of the bulk projection on the (001) surface at binding energy $E_\textrm{B} = 0.08$ eV, Fig. \ref{Fig2}B. The size of the surface Brillouin zone matches and shows similar features, including (1) a square feature around $\bar{\Gamma}$, (2) a wedge-like feature along the $\bar{\Gamma}-\bar{X}$ lines and (3) a large rounded feature along the $\bar{X}-\bar{M}$ lines. We can better understand these states by considering the line nodes calculated in Fig. \ref{Fig1}E and plotting their projection on the (001) surface, Fig. \ref{Fig2}C. By matching up the line node projection with the calculated and measured constant energy surfaces, we see that all key features in our spectra appear to be closely related to the line nodes. However, to directly demonstrate the line node, we should observe a band crossing on an $E_\textrm{B}-k_{||}$ cut passing through a line node. We consider an $E_\textrm{B}-k_x$ cut in calculation at $k_y = -0.4$ $\textrm{\AA}^{-1}$ and we readily pinpoint the cones associated with the blue and yellow line nodes, Fig. \ref{Fig2}D. However, we note that the match between Figs. \ref{Fig2}A,B suggests the sample is hole-doped by $\sim 0.08$ eV. From Fig. \ref{Fig2}D, the blue line node (white arrow) is still below the Fermi level under a $\sim 0.08$ eV hole doping. Based on this analysis, we focus on the blue line node.

We now directly pinpoint a line node in \s\ by ARPES. We emphasize that because each line node individually is associated with its $\pi$ Berry phase invariant, the observation of any number of line nodes demonstrates a topological phase. In the following discussion we focus on demonstrating one line node, leaving for the Supplementary Materials an exhaustive experimental and theoretical discussion of the complex line node network in this system. We present a series of $E_\textrm{B}-k_{||}$ spectra cutting through the blue line node, centered at different values of $k_y$, Fig. \ref{Fig2}E-H. We directly observe a band crossing, indicated by the green arrows. We note that the behavior of the band crossing as a function of $k_y$ is unusual compared to ordinary Dirac or Weyl semimetals. In particular, Dirac and Weyl semimetals host point touchings of bands, so a spectrum will show a gapless crossing on an $E_\textrm{B}-k_x$ cut at some special $k_y = k_\textrm{P}$. Away from $k_\textrm{P}$, the cone gaps out. By contrast, here we observe a cone in a range of $k_y$, as marked by the green arrows, showing an extended one-dimensional band crossing. This is the key experimental signature of a line node. We also find that the line node disperses downward in energy by $\sim 0.02$ eV as we move away from $\bar{\Gamma}$, as allowed by symmetry and consistent with our calculation. Next, we find that beyond the edge of the line node the cone gaps out, Fig. \ref{Fig2}I. This is expected quite generally for a line node in the sense that if we study an $E_\textrm{B}-k_x$ cut as function of $k_y$, the cut will exhibit a cone until we move off of the line node entirely. When $k_y$ moves past the line node, the cone will gap out. Quantitatively, we see that the cone gaps at $k_y > -0.5$ $\textrm{\AA}^{-1}$, exactly where we expect the extremum of the line node in theory. Lastly, we find an excellent quantitative match in the line node dispersion between experiment and theory, provided we take into account the $\sim 0.08$ eV hole doping in our sample.

We can also identify the line node by its characteristic evolution on a constant energy surface as a function of $E_\textrm{B}$. As we move from $E_\textrm{F}$ to deeper $E_\textrm{B}$ we see that the line node changes from a $<$ shape to a point to a $>$ shape, see blue dotted lines in Figs. \ref{Fig3}A-D. We observe a consistent evolution in the calculation, Figs. \ref{Fig3}E,F. We can understand this evolution by considering a generic line node with some energy dispersion, Figs. \ref{Fig3}G,H. For energies above the line node, the constant energy surface intersects only the upper cone of the line node, giving {\bf I}, as in Fig. \ref{Fig3}A. For energies which cross the line node, we find electron and hole pockets intersecting at a point, as in {\bf II}, Fig. \ref{Fig3}C. As we continue to move downward in $E_\textrm{B}$, the intersection point traces out the line node, shifting from left to right. As a side note, while we can trace out the line node in this way in our ARPES spectra, it is difficult at any given $E_\textrm{B}$ to pinpoint both the electron and hole pockets on either side of the intersection point, possibly because the photoemission cross-section is dominated by the intersection point for this range of $E_\textrm{B}$. As is clear from Fig. \ref{Fig3}A-D, we see either the electron pocket, a point or the hole pocket. Lastly, as we scan $E_\textrm{B}$ below the line node, the intersection point zips closed the entire electron pocket and zips open the entire hole pocket, as in {\bf III}, Fig. \ref{Fig3}D. In this way, a study of the $E_\textrm{B}$ dependence of the constant energy surface in ARPES again demonstrates a line node in \s.


Lastly, we observe drumhead surface states in Co$_2$MnGa, connecting to the yellow line nodes. We have not discussed the yellow line nodes in our data so far, but the analysis follows the same general outline as for the blue line node, so we leave this discussion to the Supplementary Materials, Sect. 2 and Figs. S10, S11. To demonstrate a drumhead, we consider $E_\textrm{B}-k_{||}$ ARPES cuts taken on a diagonal shifted off $\bar{\Gamma}$, Fig. \ref{FigDrum}A-C, where the momentum-space location is fixed from cut to cut but where we have changed the photon energy of the measurement. We observe a cone feature (marked by the yellow lines on Fig. \ref{FigDrum}D), two $U$-shaped bands emanating out from the cone (green lines), and two additional outer cone features (unmarked). These results are consistent with an analogous cut in calculation, Fig. \ref{FigDrum}E. Based on our analysis of the yellow line nodes, again see the Supplementary Materials, we argue that the three cone features correspond to three yellow line nodes, which we mark by the orange arrows in Figs. \ref{FigDrum}D-F. Next, we argue that the $U$-bands form a drumhead surface state. In our ARPES spectra, we observe that the $U$-bands are pinned to the crossing point of the center yellow line node and then extend outward and terminate at the outer yellow line nodes. Moreover, in calculation, we observe a similar band structure and we directly see that the $U$-band is a surface state. To check experimentally if the $U$-bands are surface states, we carry out a photon energy dependence, Fig. \ref{FigDrum}G. We observe no dispersion from $h \nu = 34$ eV to $48$ eV, as marked by the vertical black line, see also Supplementary Materials Fig. S14 and the accompanying discussion. This result suggests that the $U$-bands are surface states. Lastly, we observe the in-plane dispersion of the $U$-bands, see Supplementary Materials Fig. S15, and we consistently find that they terminate on the outer yellow line nodes. These observations suggest that the $U$-bands form a drumhead surface state stretching across the outer yellow line node. We note that our result represents the first observation of a drumhead surface state in any line node semimetal, magnetic or non-magnetic. Also, we point out that it is perhaps unexpected that the surface state is pinned by the center yellow line node, because this line node projects on its side, not on its face. Na\"ively, we would expect that the $\pi$ Berry phase of this line node protects no surface states and, consequently, the drumhead should not be pinned, but should pass through the yellow line node generically. In the Supplementary Materials, we elaborate on some open theoretical questions raised by our work regarding drumhead surface states.


Our ARPES spectra show a line node in \s, demonstrating the first three-dimensional topological magnetic phase and the first topological magnetic metal in any dimension. We argue that our results may open a new direction of research studying the interplay of topology and magnetism. While few topological magnets have so far been demonstrated, it has long been known theoretically that systems without time-reversal symmetry also host a rich topological classification \cite{schnyder}. If we further consider crystal symmetries, more phases arise, including (1) the magnetic Weyl semimetal \cite{pyrochlore, spinel}, Fig. \ref{Fig4}A; (2) the antiferromagnetic topological insulator, where the magnetic order breaks time-reversal symmetry but respects a non-symmorphic symmetry consisting of time-reversal followed by a lattice translation \cite{JoelMoore}, Fig. \ref{Fig4}B; (3) the magnetic line node observed in this work, Fig. \ref{Fig4}C; and (4) higher Chern number magnetic Weyl points protected by crystal rotation symmetry \cite{chen}, Fig. \ref{Fig4}D. Quite generally, we expect that different kinds of magnetic order will allow different topological phases. Also, the magnetic transition temperature will be naturally associated with a topological phase transition. Finally, band inversions need not be driven by spin-orbit coupling, but rather by magnetic spin-splitting or a related magnetic parameter. On the other hand, the spin-orbit coupling, as well as the dipole-dipole interaction, will determine the coupling of the topological objects to the direction of magnetization. As mentioned in the introduction, this may allow the topological invariants to flip by changing the direction of magnetization of a sample. Indeed, it is already known that thin films of Co$_2$MnGa exhibit a tunable magnetic anisotropy as a function of thin film thickness \cite{anisotropy}. This tunable anisotropy could be used to change the shapes of the line nodes or dissolve them into Weyl points, an effect that could be directly observed in ARPES. By demonstrating a magnetic topological phase we open the way to studying these fascinating phenomena in Co$_2$MnGa and other magnets.



We discuss two possible transport phenomena which are motivated by our work and accessible in Co$_2$MnGa. (1) It was recently proposed that line node semimetals may give rise to a novel Hall-like current arising from the parity anomaly of quantum field theory \cite{parityanomaly}. Crucially, the proposal considers a class of line nodes respecting parity, $P$, and time-reversal, $T$, symmetry, while Co$_2$MnGa breaks $T$ symmetry. Therefore, our work motivates a theoretical study of anomalous currents from the parity anomaly in the case of line node semimetals which break $P$ or $T$. Such currents could be observed experimentally in Co$_2$MnGa. (2) Our work suggests a new approach to generating spin-polarized currents for spintronics or other applications. A common approach uses a half-metal, which naturally gives rise to longitudinal spin-polarized currents. Indeed, among the cobalt-based full Heuslers, Co$_2$MnSi has attracted more interest from the point of view of spintronics than Co$_2$MnGa in recent years, in part because Co$_2$MnSi is a half metal, while Co$_2$MnGa has a large pocket in the minority spin at $\Gamma$. Therefore, it is suggestive that the transverse, anomalous Hall current may be spin-polarized in Co$_2$MnGa, even though it is not a half metal, which is due to the fact that we observe in our work that the line nodes only arise in the majority spin bands in Co$_2$MnGa, and these line nodes concentrate most of the Berry curvature in the band structure. Since the Berry curvature provides the intrinsic contribution to the anomalous Hall conductivity, the anomalous Hall current is mostly associated with majority spin bands. Equivalently, since all line nodes consist of majority spin states, we expect the anomalous Hall current to be highly spin-polarized. In fact, an earlier theoretical study of the anomalous Hall response of cobalt-based Heuslers predicted an unexpectedly spin-polarized Hall current, even in cases where the band structure was not half-metallic and, indeed, specifically in the case of Co$_2$MnGa \cite{spinAHE}. This may allow the development of an anomalous Hall spin filter. Our observation of topological line nodes in Co$_2$MnGa  opens a rich direction of research studying the interplay of topology with magnetism.




\section{Methods}

\textit{Crystal growth}: Single crystals of Co$_2$MnGa were grown using the Bridgman-Stockbarger crystal growth technique. First, we prepared a polycrystalline ingot using the induction melt technique with the stoichiometric mixture of Co, Mn and Ga metal pieces of 99.99\% purity. Then, we poured the powdered material into an alumina crucible and sealed it in a tantalum tube. The growth temperature was controlled with a thermocouple attached to the bottom of the crucible. For the heating cycle, the entire material was melted above 1200$^{\circ}$C and then slowly cooled below 900$^{\circ}$C. A photograph of the grown crystal is displayed in the Supplementary Materials, Fig. S1A. We analyzed the crystals with white beam backscattering Laue X-ray diffraction at room temperature. The samples show very sharp spots, that can be indexed by a single pattern, suggesting excellent quality of the grown crystals without any twining or domains. We show a representative Laue diffraction pattern of the grown Co$_2$MnGa crystal superimposed on a theoretically-simulated pattern, Fig. S1B. The crystal structure is found to be cubic $Fm\bar{3}m$ with lattice parameter $a=5.771(5)$ $\textrm{\AA}$. We present magnetic characterization of our Co$_2$MnGa samples in the Supplementary Materials, Section 1.\\

\textit{Angle-resolved photoemission spectroscopy}: ARPES measurements were carried out at Beamlines 5-2 and 5-4 of the Stanford Synchrotron Radiation Lightsource, SLAC in Menlo Park, CA, USA with a Scienta R4000 electron analyzer. The angular resolution was better than 0.2$^{\circ}$ and the energy resolution better than 20meV, with a beam spot size of about $50 \times 40$ $\mu$m for Beamline 5-2 and $100 \times 80$ $\mu$m for Beamline 5-4. Samples were cleaved $\textit{in situ}$ and measured under vacuum better than $5 \times 10^{-11}$ Torr at temperatures $<$ 25K.\\

\textit{First-principles calculations}: Numerical calculations of Co$_{2}$MnGa were performed within the density functional theory (DFT) framework using the projector augmented wave method  as implemented in the VASP package \cite{DFT1, DFT2,DFT3}. The generalized gradient approximation (GGA) \cite{DFT4} and a $\Gamma$-centered $k$-point $12 \times 12 \times 12$ mesh were used. Ga $s, p$ orbitals and Mn, Co $d$ orbitals were used to generate a real space tight-binding model, the Wannier function. The surface states on a (001) semi-infinite slab were calculated from the Wannier function by an iterative Green's function method.

\section{Acknowledgments}

I.B. acknowledges the support of the US National Science Foundation GRFP and the Harold W. Dodds Fellowship of Princeton University. Work at Princeton University is supported by the Emergent Phenomena in Quantum Systems (EPiQS) Initiative of the Gordon and Betty Moore Foundation under Grant No. GBMF4547 (M.Z.H.) and by the National Science Foundation, Division of Materials Research, under Grants No. NSF-DMR-1507585 and No. NSF-DMR-1006492. H.L. acknowledges the support of the Singapore National Research Foundation under NRF Award No. NRF-NRFF2013-03. Use of the Stanford Synchrotron Radiation Lightsource (SSRL), SLAC National Accelerator Laboratory, is supported by the US Department of Energy, Office of Science, Office of Basic Energy Sciences under Contract No. DE-AC02-76SF00515. The authors thank Donghui Lu and Makoto Hashimoto for enthusiastic beamtime support at SSRL Beamlines 5-2 and 5-4.


\clearpage
\begin{figure}
\centering
\includegraphics[width=12cm,trim={2in 5.5in 2in 1.3in},clip]{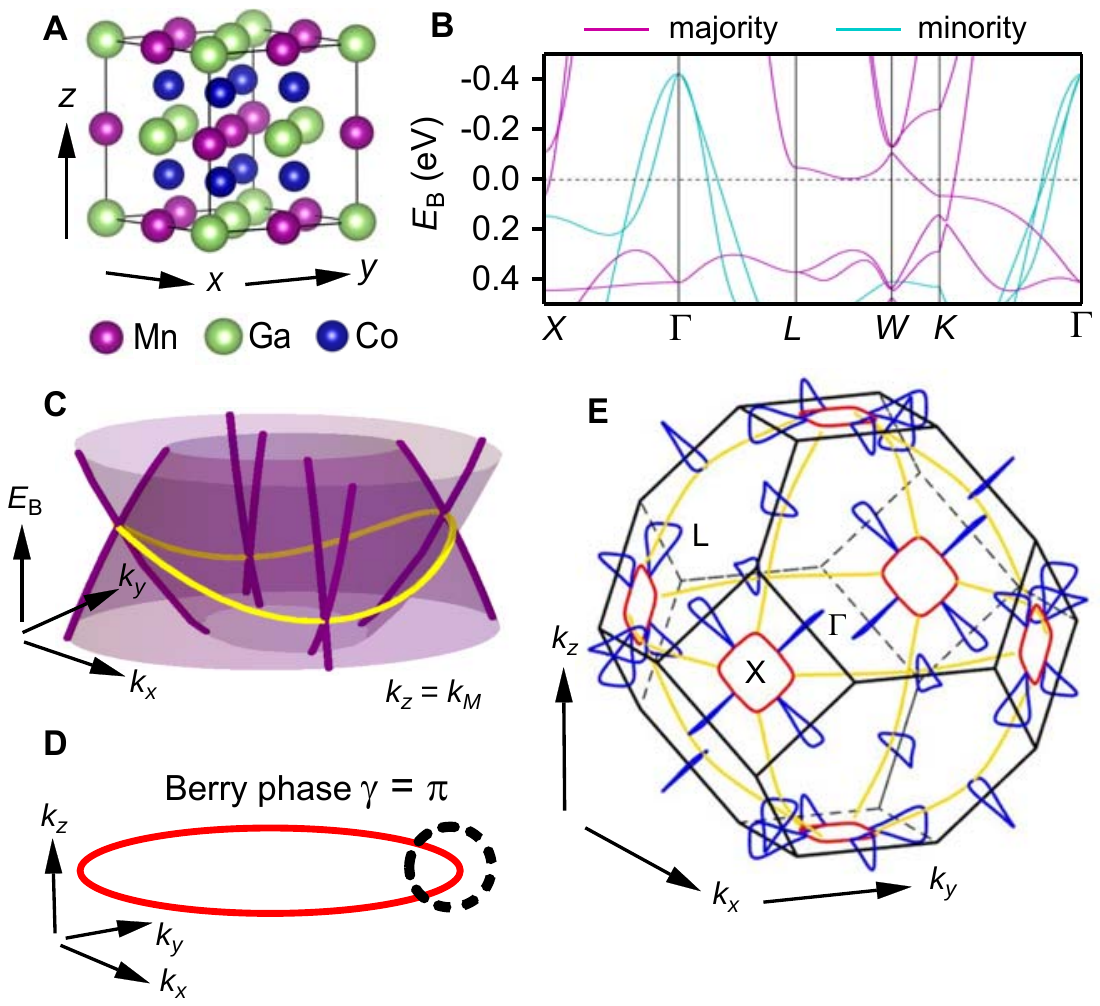}

\caption{\label{Fig1} {\bf Prediction of a magnetic line node in \s.} {\bf A}, Crystal structure of \s, the full Heusler crystal structure. {\bf B}, Bulk band structure of \s\ along high-symmetry lines in the ferromagnetic phase. The majority spin shows a band crossing between a single valence and conduction band near $E_\textrm{F}$. {\bf C}, A line node (yellow curve) is a band degeneracy along an entire closed curve in the Brillouin zone. Such a degeneracy can be protected by a mirror symmetry of the crystal structure, in which case the line node must live in the mirror plane of the Brillouin zone, labelled here without loss of generality as $k_z = k_\textrm{M}$. However, the line node is quite generally allowed to disperse in energy, as illustrated. Any $E_\textrm{B}-k_{||}$ cut through the line node shows a cone-shaped dispersion (purple surfaces). {\bf D}, A line node is an intrinsically topological object, with a $\pi$ Berry phase accumulated on any closed loop encircling the line node. {\bf E}, Plot of twofold degenerate line nodes in \s\ between the majority spin conduction and valence band, from \textit{ab initio} calculation. It is perhaps to be expected that line nodes proliferate in \s\ because the crystal structure has many mirror planes. The line nodes within each family (red, blue or yellow) are copied into each other by the symmetries of the crystal lattice.}
\end{figure}

\clearpage
\begin{figure}
\centering
\includegraphics[width=16cm,trim={1in 5.8in 1in 1in},clip]{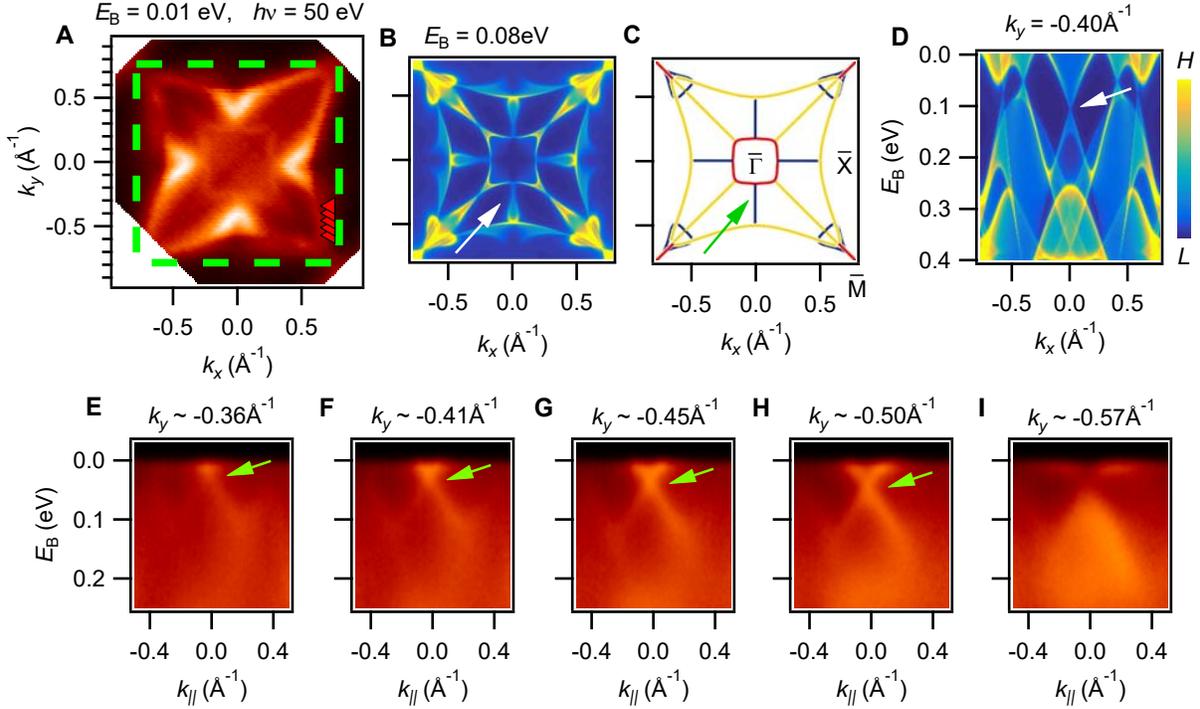}

\caption{\label{Fig2} {\bf Observation of a magnetic line node in ARPES in \s.} {\bf A}, Fermi surface measured at incident photon energy $h\nu = 50$ eV, with surface Brillouin zone marked (green box). {\bf B}, \textit{Ab initio} constant energy surface at binding energy $E_\textrm{B} = 0.08$ eV below $E_\textrm{F}$, on the (001) surface with MnGa termination, showing qualitative agreement with the ARPES. We note that this suggests a hole-doping of $\sim 0.08$ eV in our samples. The colors indicate the spectral weight of a given state on the surface, see the color bar. {\bf C}, Projection of the line nodes calculated in Fig. \ref{Fig1}E on the (001) surface. A comparison with the measured and calculated constant energy surface suggests that the features in the band structure all arise from the proliferation of line nodes. {\bf D}, \textit{Ab initio} $E_\textrm{B}-k_x$ cut through the blue and yellow line nodes. We focus on the blue line node (white arrow in {\bf D}, green arrow in {\bf C}) which we predict remains well below $E_\textrm{F}$ even under a hole doping of $\sim 0.08$ eV. {\bf E-H}, $E_\textrm{B}-k_{||}$ cuts through the blue line node centered at different $k_y$, indicated by the red arrows in {\bf A}. We observe crossings in a range of $k_y$, showing an extended one-dimensional band crossing. {\bf I}, For $k_y \sim -0.5$ $\textrm{\AA}^{-1}$ we move past the edge of the blue line node and the cone gaps out, as expected. These ARPES spectra are the smoking gun signature of a magnetic line node in \s.}
\end{figure}

\clearpage
\begin{figure}
\centering
\includegraphics[width=16cm,trim={1in 4.3in 1in 1in},clip]{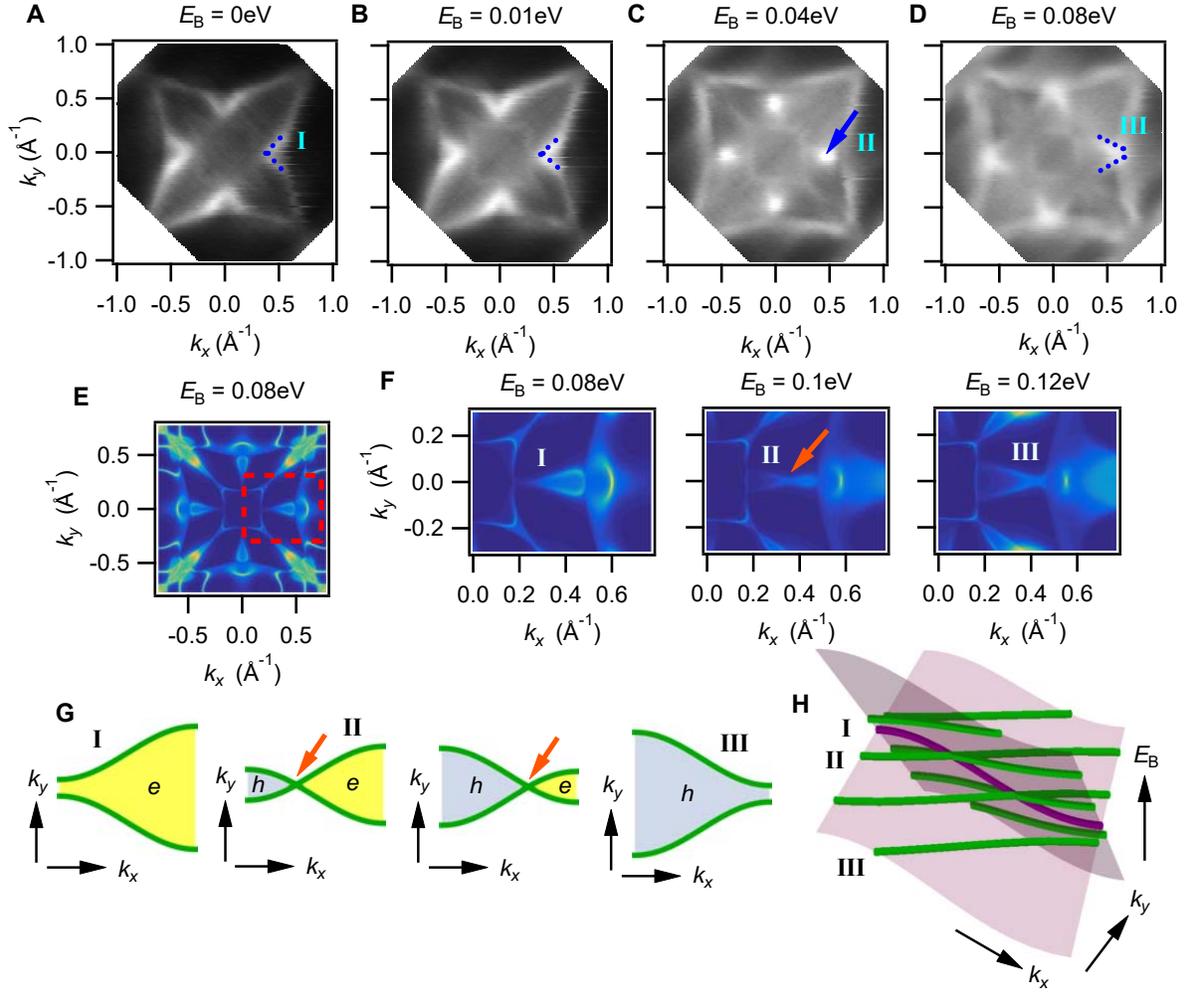}

\caption{\label{Fig3} {\bf Dispersive evolution of the line node in energy in \s.} {\bf A-D}, Constant energy surfaces measured by ARPES at different $E_\textrm{B}$, showing the evolution of the valence and conduction bands of the line node. {\bf E}, \textit{Ab initio} Constant energy surface with red boxed region marking {\bf F}, the zoomed-in constant energy surfaces at different $E_\textrm{B}$. {\bf G}, {\bf H}, Cartoon of the constant energy surfaces of a generic line node. As we scan the energy from shallower to deeper $E_\textrm{B}$, the constant-energy surface evolves from {\bf I}, an electron pocket, to {\bf II}, a hole and electron pocket touching at a point (blue arrow in {\bf C}, red arrows in {\bf F} and {\bf G}), to {\bf III}, a hole pocket. The touching point at each energy is a point on the line node. As we scan downward through $E_\textrm{B}$, the intersection point zips the electron pock away and unzips the hole pocket. We can directly observe this evolution in our ARPES spectra and our \textit{ab initio} calculation. This provides an independent demonstration of a topological magnetic line node in \s.}
\end{figure}

\clearpage
\begin{figure}
\centering
\includegraphics[width=16cm,trim={1in 6.5in 1in 1in},clip]{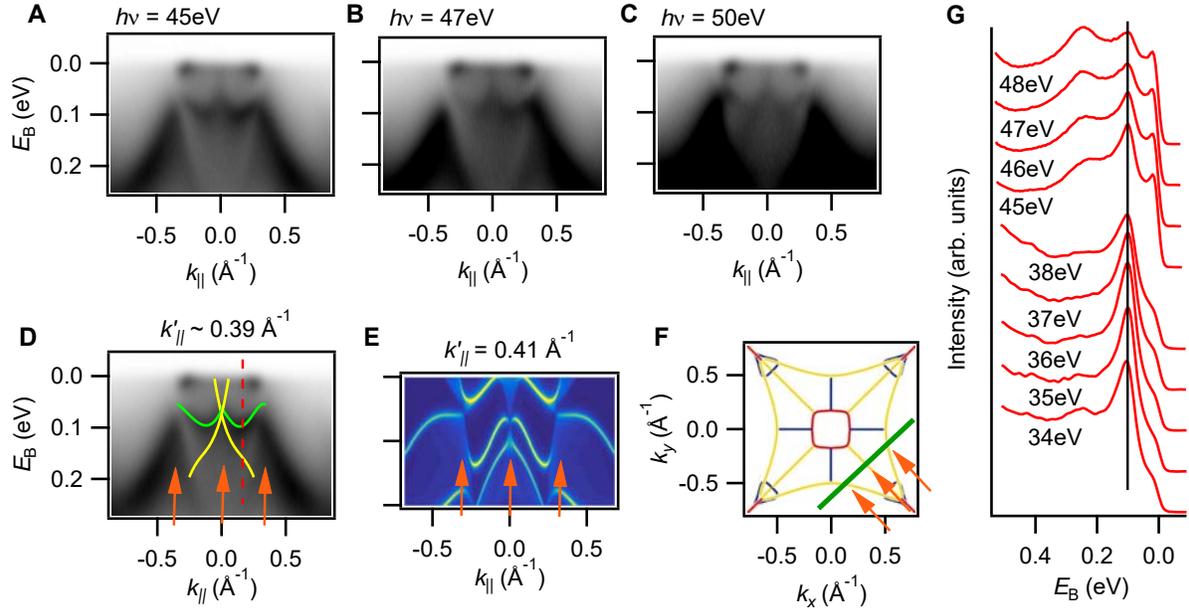}
\caption{\label{FigDrum} \textbf{Observation of drumhead surface states in Co$_2$MnGa.} \textbf{A}-\textbf{C}, ARPES spectra showing drumhead surface states at different photon energies. \textbf{D}, Same as \textbf{A} but with key features marked by hand drawn guides: center yellow line node (yellow marking), drumhead surface state (green marking). The drumhead stretches across the outer yellow line node (outer orange arrows) and is pinned by the center yellow line node (center orange arrow). \textbf{E}, An analogous cut from \textit{ab initio} calculation, consistent with experiment and, in particular, also showing a drumhead surface state. \textbf{F}, Same as Fig. \ref{Fig2}C but with a green line marking the location of the ARPES spectra in \textbf{A}-\textbf{C}. Again, the orange arrows point to the three yellow line nodes crossing the cut. \textbf{G}, Photon energy dependence of an energy distribution curve (EDC) passing through the drumhead (red dotted line in \textbf{D}). The peaks marked by the black vertical line correspond to the drumhead surface state. We observe no dispersion as a function of photon energy, providing direct evidence that the peaks correspond to a surface state. In this way, our photon energy dependence also suggests that the band of interest is a drumhead surface state. We note that we have observed the first drumhead surface state in any material, magnetic or non-magnetic.}
\end{figure}

\clearpage
\begin{figure}
\centering
\includegraphics[width=16cm,trim={1in 5.2in 1in 1in},clip]{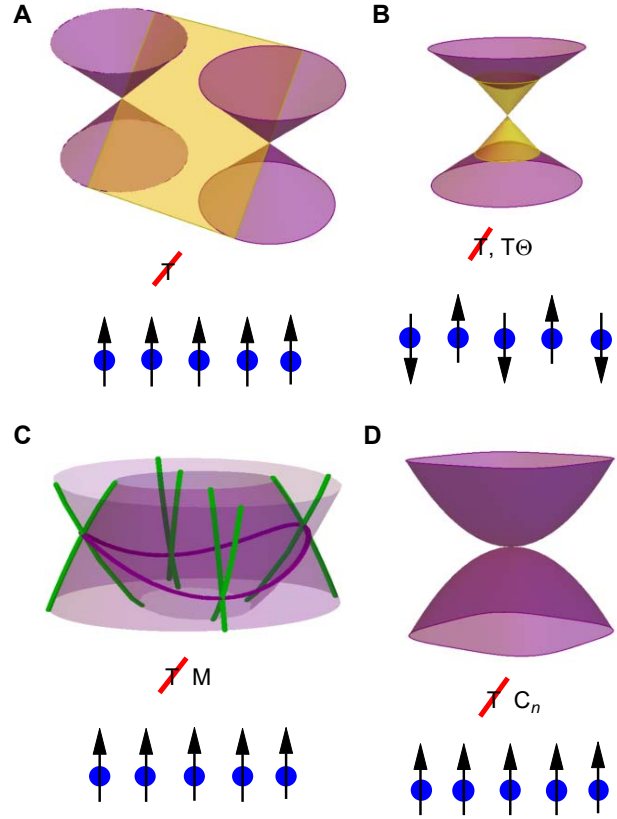}

\caption{\label{Fig4} {\bf Natural magnetic topological groundstates.} Various time-reversal symmetry breaking topological phases have been proposed, including {\bf A}, the well-known magnetic Weyl semimetal \cite{pyrochlore, spinel}; {\bf B}, the antiferromagnetic topological insulator \cite{JoelMoore}; {\bf C}, the line node realized in this work; and {\bf D}, higher Chern number Weyl points protected by magnetic space group symmetries \cite{chen}. Our work opens the way to the discovery of these and other magnetic topological phases, along with phenomena arising from fascinating interplay between topological invariants and magnetism.}
\end{figure}

\clearpage
\begin{center}
\textbf{\large Supplementary Materials: A three-dimensional magnetic topological phase}
\end{center}

\setcounter{equation}{0}
\setcounter{figure}{0}
\setcounter{table}{0}
\makeatletter
\renewcommand{\theequation}{S\arabic{equation}}
\renewcommand{\thefigure}{S\arabic{figure}}
\renewcommand{\thetable}{S\arabic{table}}

In this Supplementary Materials, we first characterize the magnetic properties of our Co$_2$MnGa samples. Then, we provide a systematic discussion of the rich network of line nodes in Co$_2$MnGa and present additional ARPES and calculation data omitted from the main text. Lastly, we comment on the relationship between the topological invariant and drumhead surface states in line node semimetals in general, highlighting some theoretical questions which remain unaddressed.\\

\textbf{\large{1. Magnetic \& XPS characterization of Co$_2$MnGa}}\\

We present a measurement of the DC magnetization, carried out using a Quantum Design vibrating sample magnetometer (VSM) with temperature range $2-400$ K and magnetic field up to $7$ T. The transport measurements were performed using a Quantum Design physical property measurement system with low-frequency alternating current. The temperature-dependent field-cooled (FC) and zero-field-cooled (ZFC) magnetization of the Co$_2$MnGa crystal were recorded in a DC magnetic field of 200 Oe applied along the [001] direction in the temperature range of $300-1000$ K, Fig. \ref{FigS2}A. The magnetization falls sharply above 660 K resulting in a ferromagnetic to paramagnetic transition with Curie temperature $T_\textrm{C} \sim 687$ K, in good agreement with previously published results \cite{CMGCurie}. The magnetic hysteresis loop recorded at 2 K shows a soft ferromagnetic behavior, \ref{FigS2}b. The magnetization saturates above $\sim 0.5$ T field with saturation magnetization $M_\textrm{S} \sim 4.0 \mu_{\textrm{B}}$/f.u. Evidently, the compound follows the Slater-Pauling rule, $M_\textrm{S} = N-24$, where $N$ is the number of valance electrons, $N = 28$ for Co$_2$MnGa, suggesting half-metallic behavior. We see a ferromagnetic loop opening with coercive field $\sim 35$ Oe, Fig. \ref{FigS2}B, inset. Lastly, we study the temperature dependent longitudinal resistivity of our samples $\rho_{xx} (T) $, with zero applied magnetic field, Fig. \ref{FigS3}. We applied current along the [100] direction. Clearly the compound shows metallic behavior throughout the temperature range with very low residual resistivity: $\rho_{xx}$ $(2\textrm{K}) \sim 5.6 \times 10^{-5}$ $\Omega$ $\textrm{cm}$. The residual resistivity ratio (RRR) was found to be $\rho_{xx}$ $(300 \textrm{K})/\rho_{xx}$ $(2 \textrm{K}) = 2.6$ for our Co$_2$MnGa single crystals.\\

We present a core level XPS spectrum of our Co$_2$MnGa samples, Fig. \ref{FigCore}. We clearly identify the relevant core levels, including Ga $3d$, Mn $3p$ and Co $3p$ peaks, suggesting a high quality sample.\\

\textbf{\large{2. A network of line nodes in Co$_2$MnGa}}\\

\textit{Overview of the line nodes}: In the main text we focused on the blue line node in Co$_2$MnGa. Here, we give a more systematic introduction to the full network of line nodes. Recall that a line node is a one-dimensional crossing between a pair of bands. While the line nodes are contained in the mirror planes of the bulk Brillouin zone, they are allowed to disperse in energy. As a result, it is instructive to plot the line node as a function of $k_x, k_y$ and $E_\textrm{B}$, where $k_x$ and $k_y$ without loss of generality are the two momentum axes of the mirror plane. There are three independent line nodes in Co$_2$MnGa, which we denote the red, blue and yellow line nodes. We plot the dispersions of these three line nodes from calculation in Fig. \ref{FigCalc}B-D. We note that the energies are marked with respect to the Fermi level observed in numerics. Since we find in experiment that the Fermi level is at about $-0.08$ eV relative to calculation, it cuts through all of the line nodes. Next, we can collapse the energy axis and plot the line nodes in $k_x, k_y, k_z$ space. We find that although we start with three independent line nodes, these are each copied many times within the Brillouin zone by the symmetry operations of the crystal lattice, giving rise to a complicated line node network. The final arrangment of line nodes is summarized in main text Fig. 1E, shown again here as Fig. \ref{FigCalc}A. Next, since we are studying the (001) surface, it's useful to consider how the line nodes project into the surface Brillouin zone. For instance, the center red line node will project straight up on its face around $\bar{\Gamma}$, see for instance main text Fig. 2C. By contrast, the adjacent blue line nodes will project on their side, so that in fact there will be a ``double'' cone in the surface projection. Moreover, that blue line node projection will no longer appear as a closed curve in the surface Brillouin zone. Rather, it will form a finite line node segment, as can be seen in main text Fig. 2C (green arrow). Lastly, we find yellow line nodes in the $k_x, k_y$ plane and in the $k_x,  k_z$ plane, so on the (001) surface there will be a double yellow line node along $\bar{\Gamma} - \bar{M}$ and a single yellow line node near $\bar{M} - \bar{X}$, see Fig. \ref{FigYellow1}B.\\

\textit{The Berry curvature}: Next, we calculate the integrated Berry curvature $\Omega(E)$ of Co$_2$MnGa, Fig. \ref{FigBerry}A,B. We observe a peak at the approximate energy of the line nodes, showing that the Berry curvature is dominated by the contribution from the line nodes. This suggests that the anomalous Hall response of the system arises mostly from the line nodes. We also note that the line nodes are essentially magnetic in the sense that they are removed if we remove the magnetic order. Indeed, the band structure without magnetic order changes drastically, Fig. \ref{FigNonMag}. At the same time, we note that a new family of line nodes will likely arise in the non-magnetic band structure.\\

In the rest of this supplementary section, we present ARPES data on each of these line nodes in turn. We find that the blue line node node is most readily observed in our spectra with rich structure, as already discussed at length in the main text. We can also demonstrate the yellow line node, although we find that the spectral intensity is weaker, so we must rely on a careful analysis of momentum distribution curves (MDCs) and on a comparison with calculations. Lastly, we can observe clear signatures of the red line node in our data, but we cannot pinpoint its band crossing.\\

\textit{Additional systematics on the blue line node}: Next, we present some additional ARPES data on the blue line node. We present complete systematics on the constant energy surface, Fig. \ref{FigBlueFS}A-I. Here we can track the evolution of the line node more carefully in $E_\textrm{B}$. Notably, we can see that the hole pocket, shown in maintext Fig. 3D, becomes larger with deeper $E_\textrm{B}$, Fig. \ref{FigBlueFS}F-I. We find that the crossing point moves away from $\bar{\Gamma}$ with deeper $E_\textrm{B}$. For instance, when cutting near the center of the line node, we find that the crossing point peak shifts from $k_x \sim 0.45 \textrm{\AA}^{-1}$, Fig. \ref{FigBlueFS}D, to $k_x \sim 0.46 \textrm{\AA}^{-1}$, Fig. \ref{FigBlueFS}E (distribution curve analysis not shown). This analysis further supports our observation of a line node.\\

To provide another perspective on our data, we can cut parallel to the blue line node, Fig. \ref{FigBluePar}A-I, with locations of the cuts shown in Fig. \ref{FigBluePar}J. In contrast to main text Fig. 2, where we cut perpendicular to the line node, here we cut \textit{along} the line node, so that we will encounter an $E_\textrm{B}-k_x$ cut which contains the entire line node, see Fig. \ref{FigBluePar}K, L. As we sweep through the line node in this way, we see the conduction and valence bands approach, Fig. \ref{FigBluePar}A-D, touch each other at fixed $k_y \sim 0 \textrm{\AA}^{-1}$ along a finite range of $k_x$, Fig. \ref{FigBluePar}E, and then move apart again, Fig. \ref{FigBluePar}F-I. Note that in our data we in fact observe the line node slightly away from $k_y = 0 \textrm{\AA}^{-1}$, Fig. \ref{FigBluePar}E, probably due to a small error in the calibration of the momentum axes or a slight misalignment of the sample. Our parallel $E_\textrm{B} - k_x$ cuts again demonstrate the blue line node.\\

We can also carry out a numerical peak fitting of our ARPES spectra. We begin with the same $E_\textrm{B}-k_x$ cuts discussed in main text Fig. 2 and we choose by hand the energy distribution curve (EDC) passing through the center of the cone, as marked by the yellow arrows, Fig. \ref{FigBlueFit}A-E. We fit these EDCs to a function of the following form,

\begin{equation*}
I(x) = (C + L_1(x) + L_2(x))f(x)
\end{equation*}

\begin{equation*}
L_i(x) = \frac{A_i^2}{(x - B_i)^2 + C_i^2}  \hspace{1cm}    f(x) = (\exp(\beta(x - \mu)) + 1)^{-1}
\end{equation*}
\\ 
We include two Lorentzian peaks $L_1(x)$ and $L_2(x)$ for Fig. \ref{FigBlueFit}A-D, where one peak, LN corresponds to the line node crossing, while the deeper peak corresponds to an irrelevant valence band VB' which is useful for improving the fit. We also include the Fermi-Dirac distribution $f(x)$ and a constant offset $C$ which we find empirically provides a good fit and which we interpret as a background spectral weight approximately constant within the energy range of the fit. We find an excellent fit using only one LN peak, suggesting that we observe a true band crossing. Next, for Fig. \ref{FigBlueFit}E, we fit three peaks because we have $k_y > 0.5 \textrm{\AA}^{-1}$ so we have passed the end of the projected line node and the bands gap out, see discussion in the main text. This gives VB and CB, the peaks corresponding to the conduction and valence bands of the line node. Again this result is consistent with calculation and supports our earlier interpretation of our spectra. We note that the fit fails slightly for LN for Fig. \ref{FigBlueFit}I, an interesting discrepancy. This mismatch is perhaps to be expected because we are right at the edge of the line node at $k_y = 0.5 \textrm{\AA}^{-1}$ so we may be observing a very small band gap smeared out by the Lorentzian broadening. Another interesting explanation considers the detailed dispersion of the blue line node. Specifically, from Fig. \ref{FigCalc}C we find a sharp upward dispersion at the extremum of the line node projection, see also Fig. \ref{FigBlueTrack}. Due to broadening along $k_y$, we may expect a plateau structure in the EDC, because we capture LN peaks from a range of $k_y$ which disperse sharply in $E_\textrm{B}$. By fitting the peaks to Lorentzians, we provide further evidence that we observe a true extended one-dimensional band crossing.\\

Lastly, we take the results of our peak fitting and compare them with the calculated blue line node dispersion. We plot the LN peak maxima and the standard deviation of the peak positions, Fig. \ref{FigBlueTrack}. We estimate an error of $\sim 0.005$ eV by eye from the $E_\textrm{B}-k_y$ cuts of the line node. We ignore EDCs at $k_y > 0.45 \textrm{\AA}^{-1}$ because the plateau shape in the EDC is poorly described by a single Lorentzian, as discussed above. We compare these numerical fitting results with a first-principles calculation of the blue line node dispersion, shifted by $0.08$ eV based on our earlier determination of the doping by comparing measured and calculated constant energy surfaces, see main text Fig. 2. We find a reasonable quantitative agreement between the numerical fit and calculated blue line node dispersion. There is some expected contribution to the error from the \textit{ab initio} calculation as well as the form that we chose for our fitting function. These results again support our observation of a line node in our ARPES spectra in Co$_2$MnGa.\\


\textit{Observation of the yellow line node}: Up until this point we have discussed only the blue line node. However, we can also weakly observe the yellow line node directly in our ARPES data. We can directly identify the line node by comparing an ARPES constant energy surface and the projected nodal lines, Fig. \ref{FigYellow1}A, B. We note that there are two different ways in which the yellow line nodes can project on the (001) surface. In particular, the four yellow line nodes along $\bar{\Gamma} - \bar{M}$ are ``standing up'', so two crossings project onto the same point in the surface Brillouin zone, similar to the blue line node we discussed above. By contrast, the outer yellow line node runs in a single large loop around the entire surface Brillouin zone. It projects ``lying down'', with single crossing projections, see also the discussion of Fig. \ref{FigCalc}. In the following we focus mostly on the double yellow line node. We study constant energy surfaces at various binding energies, Fig. \ref{FigYellow1}C-E, and we observe the same $<$ to $>$ switch that we discussed in the case of the blue line node, marked in green in Fig. \ref{FigYellow1}F-H. We can see consistent behavior in the yellow line node on an \textit{ab initio} constant energy surface, Fig. \ref{FigYellow1}I-K. Note crucially that the electron to hole pocket swap as a function of $E_\textrm{B}$ takes place in the same direction in the ARPES spectra and in calculation, showing that the line node dispersion has the same slope in calculation and experiment. This provides strong evidence that we have observed the yellow line node in our ARPES spectra. We can also weakly observe the cone on an $E_\textrm{B}-k_{||}$ cut. We take diagonal cuts in the surface Brillouin zone, offset from $\bar{\Gamma}$, as marked in Fig. \ref{FigYellow2}G, H. We observe the upper cone readily near the center of the cut for $k_{||}$ closest to $\bar{\Gamma}$, \ref{FigYellow2}a, corresponding to the green marking in Fig. \ref{FigYellow1}F. As we slide away from $\bar{\Gamma}$, the band crossing and lower cone become visible, \ref{FigYellow2}b. Lastly, we can still see a faint trace of the lower cone as we continue to move further from $\bar{\Gamma}$, Fig. \ref{FigYellow2}C. From each of the spectra we take a momentum distribution curve (MDC) at some energy to pinpoint the line node. We can observe twin peaks corresponding to the upper and lower cone, \ref{FigYellow2}d,f, as well as a single peak when we cut through the line node, \ref{FigYellow2}e. In this way, we directly observe the yellow line node of Co$_2$MnGa in our ARPES spectra.\\

We can also observe signatures of the single yellow line node in our ARPES data. The outer features in Fig. \ref{FigYellow2}B, C, as well as the large off-center peaks in Fig. \ref{FigYellow2}E, F, correspond well to the expected locations of the single yellow line node. The valence band further shows a cone shape. However, we note that the conduction band appears to be mismatched from the valence band, for instance near the Fermi level at $k_{||} \sim 0.25 \textrm{\AA}^{-1}$ in Fig. \ref{FigYellow2}E. This suggests that we are perhaps observing a surface state or surface resonance which partly traces out the line node, suggesting a cone without showing a crossing. This explanation is not only supported by the ARPES spectra discussed here, but also appears to be consistent with our calculations, see main text Fig. 4C. In summary, we directly demonstrate the double yellow line node and provide signatures of the single yellow line node in our ARPES spectra.\\

\textit{Signatures of the red line node}: Next, we discuss signatures of the red line node. We study constant energy surfaces at several binding energies, Fig. \ref{FigRed}F-H, and we clearly observe a square around $\bar{\Gamma}$. This corresponds well to the projection of the red line node, as indicated in Fig. \ref{FigRed}D, providing a clear signature of the red line node. However, we note that we do not see the characteristic $<$ to $>$ swap that we observed for the blue and yellow line nodes. Going further, we study a series of $E_\textrm{B}-k_x$ cuts passing through the center square feature. We see two clear branches dispersing away from $k_x = 0 \textrm{\AA}^{-1}$ as we approach the Fermi level, corresponding to the center square on the constant energy surface, Fig. \ref{FigRed}A-C. We can directly mark these features on an MDC, Fig. \ref{FigRed}E. However, we observe no cone or crossing. This may be because the other branch of the line node has low photoemission cross section under these measurement conditions or because the states have low amplitude near the surface. It may also be that we are actually observing a surface state or a resonance state which is pushed out of the bulk gap against the bulk projection. If there is one such state, then we can expect that it will be associated with one branch of the bulk line node, so that we see no crossing. Alternatively, the absence of a surface state in the red line node could be understood within a na\"ive topological theory. First, note that we do observe a drumhead surface state in the region bounded by the single, outer yellow line node, as discussed in the main text. These two observations, (1) a drumhead inside the outer yellow line node projection and (2) no drumhead inside the red line node projection, can be understood by considering the Berry phase topological invariant associated with the line nodes. In particular, the red line node and the outer yellow line node project on top of each other, so the region bounded by the red line node actually has two line nodes projecting together. If we view each line node as contributing a $\pi$ Berry phase, the result is that the two line nodes cancel each other out and we expect no surface state within the red line node projection. In the following section, we comment further on the relationship of the drumhead surface state to the topological invariants of the line nodes. Here, we only conclude that we observe clear signatures of the red line node in that we see a square feature with excellent match between experiment and calculation.\\

\textit{Observation of line nodes at $h\nu = 35$} eV: As an additional set of systematics, we repeat our measurement under a different set of experimental conditions, specifically with incident photon energy $h\nu = 35$ eV. On the constant energy surface we can pinpoint again the blue, yellow and red line nodes, Fig. \ref{Fig35eV}A-C, by comparison with the projected line nodes. We can also directly demonstrate the blue line node from our data alone by pinpointing a band crossing on a range of $k_y$, Fig. \ref{Fig35eV}D-G. We find that at $k_y \sim -0.5 \textrm{\AA}^{-1}$ the crossing gaps out, because we cut past the end of the line node projection. This result is consistent with our data presented above, at $h\nu = 50$ eV, as well as with calculation, providing additional evidence supporting our observation of a line node in Co$_2$MnGa.\\

\textit{Systematic photon energy dependence of the drumhead surface state}: We present an extended dataset relating to main text Fig. 4, showing the drumhead surface states. In main text Fig. 4, we presented an energy distribution curve (EDC) stack cutting through the drumhead surface state at different photon energies. In Fig. \ref{DH1}, we present the full $E_\textrm{B}-k_x$ cut for each photon energy included in the stack. We observe the drumhead surface state consistently at all energies (orange arrow), Fig. \ref{DH1}A-I. Additionally, we show an EDC stack at a different momentum, $k_{||} = 0.45\textrm{A}^{-1}$, which cuts not through the drumhead surface state but the yellow line node cone, Fig. \ref{DH1}J, analogous to main text Fig. 4G. When cutting through the drumhead surface state, the EDC stack showed no dispersion in the peak energy as a function of photon energy, indicating no $k_z$ dispersion and suggesting a surface state. Here, by contrast, we can observe that the peak positions shift with photon energy (blue and cyan arrows). This shift indicates that the EDC peak corresponds to a bulk state, consistent with the expectation that the yellow line node lives in the bulk.\\

\textit{In-plane dispersion of the drumhead surface state}: We briefly study the in-plane dispersion of the drumhead surface states. At $h\nu = 35$ eV, with Fermi surface as shown in Fig. \ref{DH2}A, we study a sequence of $E_\textrm{B}-k_x$ cuts scanning through the drumhead surface state, Fig. \ref{DH2}B-E. We observe that the surface state disperses slightly downward in energy as we scan away from $\bar{\Gamma}$ and that it narrows in $k_{||}$, as expected because the line node cones move together as we approach the corner of the Brillouin zone. By demonstrating that the candidate drumhead disperses in-plane but not out-of-plane, we show that it is indeed a surface state.\\

\textit{Study of the irrelevant minority spin pocket}: We briefly noted that Co$_2$MnGa has a large minority spin pocket around the $\Gamma$ point, see main text Fig. 1B. Formally this pocket projects over $\bar{\Gamma}$ and covers the topological states of interest. In our experiment, however, we noticed that this pocket does not appear under the experimental conditions that we used to study the topological band structure. It is interesting to search for this irrelevant pocket in ARPES and understand why it does not compete with the line node and drumhead states in photoemission. To address this question, we performed a photon energy dependence on a cut passing throuh $\bar{\Gamma}$. Recall that at $h\nu = 50$ eV we observed signatures of a drumhead surface state or perhaps a bulk cone associated with the red line node, as discussed previously in this Supplementary Materials, Fig. \ref{spinmin}A. On the other hand, for $h \nu > 50$ eV we find that the red line node features disappear and a large hole pocket appears at $\bar{\Gamma}$, see Fig. \ref{spinmin}B-I with (green line) the location of the cut marked on Fig. \ref{spinmin}J. To understand these results, we note that the red line node lives at the top of the bulk Brillouin zone, suggesting that $h\nu = 50$ eV corresponds to $k_z \sim \pi$. As we change photon energy, we expect to move toward $k_z = 0$ and intersect the irrelevant bulk hole pocket. The large hole pocket observed in our data may correspond to this irrelevant bulk band. This result suggests that the large irrelevant minority spin pocket does not interfere with our measurements of the line nodes because at $h\nu = 50$ eV we cut near the top of the bulk Brillouin zone in $k_z$.\\

\textbf{\large{3. Topological protection of drumhead surface states}}\\


Here we point out a few theoretical subtleties regarding topological surface states in line node semimetals. A line node can be characterized by several related topological invariants, depending on the relevant symmetries \cite{SchnyderCa3P2, BalentsTINI, SchnyderClass, HYKeePerovskites, ChenFang}. It has further become accepted that line node semimetals host drumhead surface states, which stretch across the projection of the line node in the surface Brillouin zone. A number of theoretical works present calculations of a material or model which exhibits drumhead surface states \cite{SchnyderCa3P2, BalentsTINI, BenWiederRappe, HongmingWengGraphene, XiaoHuCu3PdN, OkamotoCaAgX}. These works consistently observe in numerics that drumhead surface states emanate from line nodes and they suggest that the topological invariant protects the drumhead. However, by drawing pictures, it is clear that there is a qualitative difference in the behavior of a drumhead surface state when perturbations are applied compared to other topological surface states. Specifically, allowed perturbations can push the drumhead out of the bulk band gap and make it degenerate with bulk projected states. This is \textit{not} the case for a $\mathbb{Z}_2$ topological insulator or an ordinary Weyl semimetal. In the case of a $\mathbb{Z}_2$ topological insulator, adding allowed perturbations can distort the surface Dirac cone in many ways, but in all cases the Dirac cone crosses the bulk band gap. In the case of an ordinary Weyl semimetal, adding allowed perturbations can cause the Fermi arc to flutter in momentum space, but it will always connect the projections of the two Weyl cones. By contrast, adding perturbations to a line node semimetal can push the drumhead into the valence or conduction band entirely. In this sense, the role of the topological invariant in protecting drumhead surface states is more subtle than for surface Dirac cones or Fermi arcs.\\

We note that this concern does not arise if a particle-hole or chiral symmetry is present in the system, such as in a superconductor or a bipartite lattice with hopping only between sublattices \cite{VishwanathTurner, HYKeePerovskites}. In such cases, the drumhead surface state is locked at zero energy and no allowed perturbations can push it up or down in energy, so the drumhead is indeed robust in the expected way. However, there is no reason to believe that ordinary electron systems are in general well-approximated by any such symmetry. The unaddressed theoretical question is: for systems which \textit{lack} particle-hole or chiral symmetry, what is the relationship between the topological invariant in a line node semimetal and its drumhead surface states? On the one hand, the drumhead surface state can be pushed out of the bulk band gap. On the other hand, existing calculation results suggest that in practice surface states do \textit{tend} to emanate out of line nodes, in a manner which appears to be robust. These numerical observations suggest that there is an interesting relationship between the topological invariant of a line node and its surface states, different from that in topological insulators or Weyl semimetals. As a final remark, we note that in our experimental data and \textit{ab initio} calculations we unexpectedly observe that the drumhead of the single yellow line node is pinned to the double yellow line node, which projects on its side rather than on its face. This is unusual because na\"ively one expects the topological invariant of the line node to play a role on the area of the projection of a line node in the surface Brillouin zone. If the line node projects on its side, it covers no area, so the topological invariant should not pin any surface state. Yet, we observe such a pinning in our ARPES data and our \textit{ab initio} calculations. We leave further discussion of these fascinating questions to future work.\\

\clearpage
\begin{figure*}[h]
\centering
\includegraphics[width=13cm]{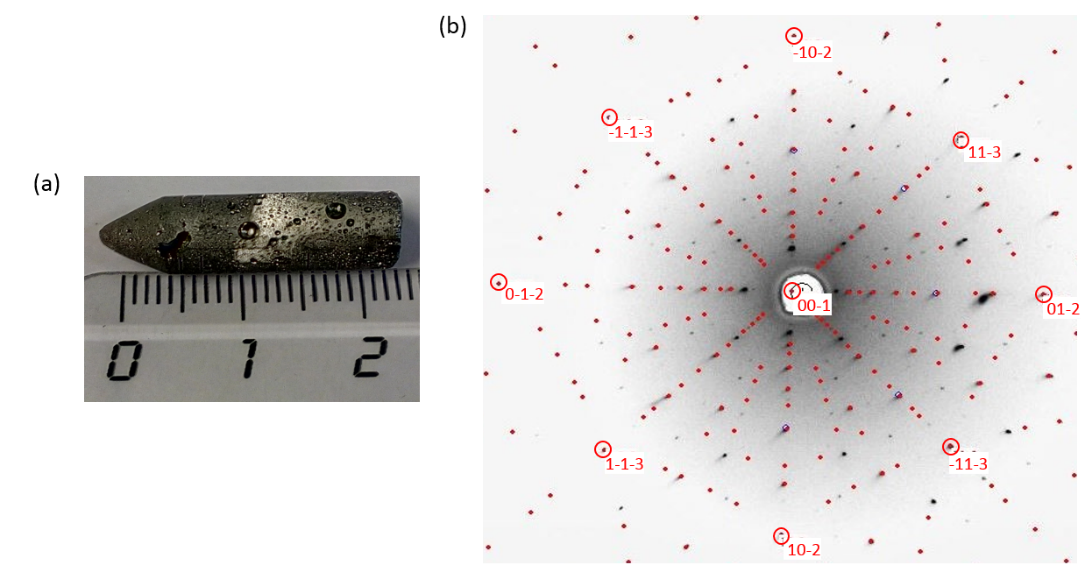}
\caption{\label{FigS1} \textbf{Crystal structure of Co$_2$MnGa.} \textbf{A}, Grown single crystal of the full Heusler material Co$_2$MnGa. \textbf{B}, Laue diffraction pattern of a [001] oriented crystal superposed with a theoretically simulated one.}
\end{figure*}

\begin{figure*}[h]
\centering
\includegraphics[width=16cm]{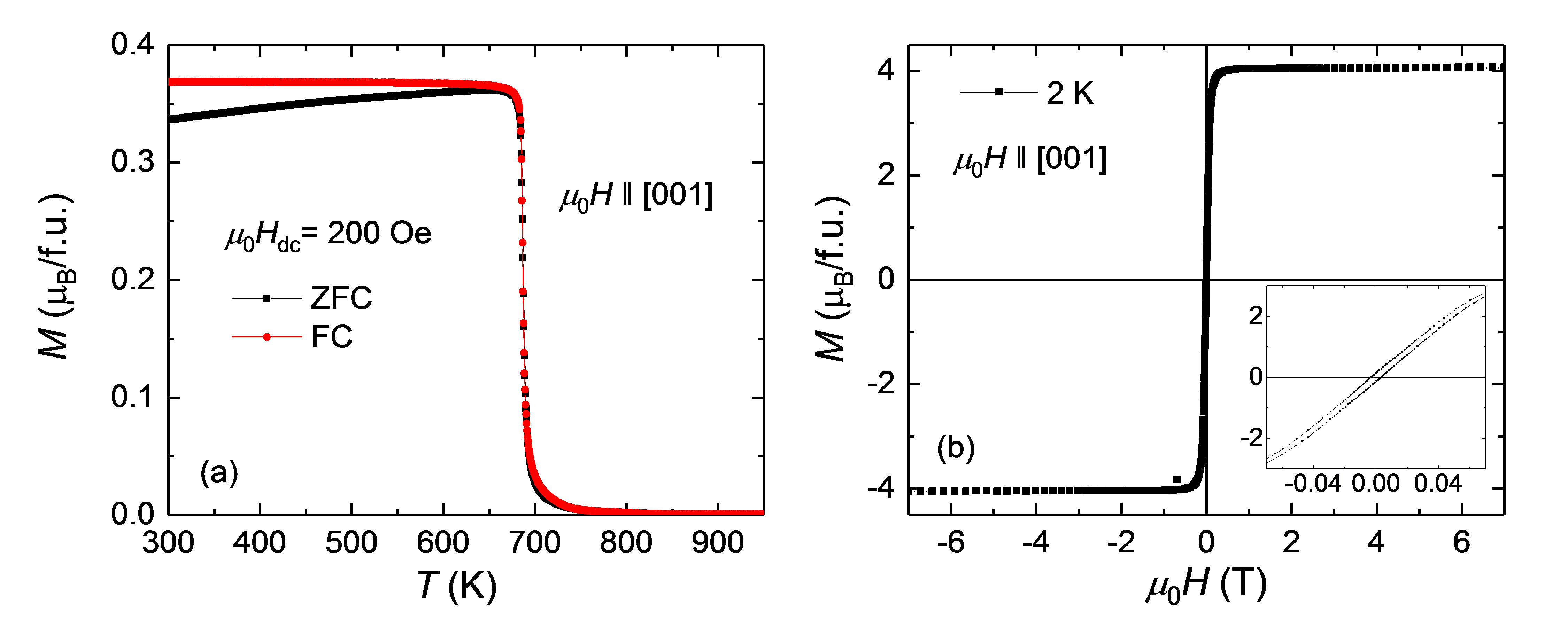}
\caption{\label{FigS2} \textbf{Curie temperature and hysteresis loop of Co$_2$MnGa.} \textbf{A}, Temperature dependent field cooled (FC) and zero field cooled (ZFC) magnetization at 200 Oe applied field for a Co$_2$MnGa single crystal. \textbf{B}, Hysteresis loop at 2 K for a [001] oriented Co$_2$MnGa crystal. Inset shows a zoomed-in view at low field, with clear hysteresis loop.}
\end{figure*}

\begin{figure*}[h]
\centering
\includegraphics[width=13cm,trim={1in 0.5in 1in 1in},clip]{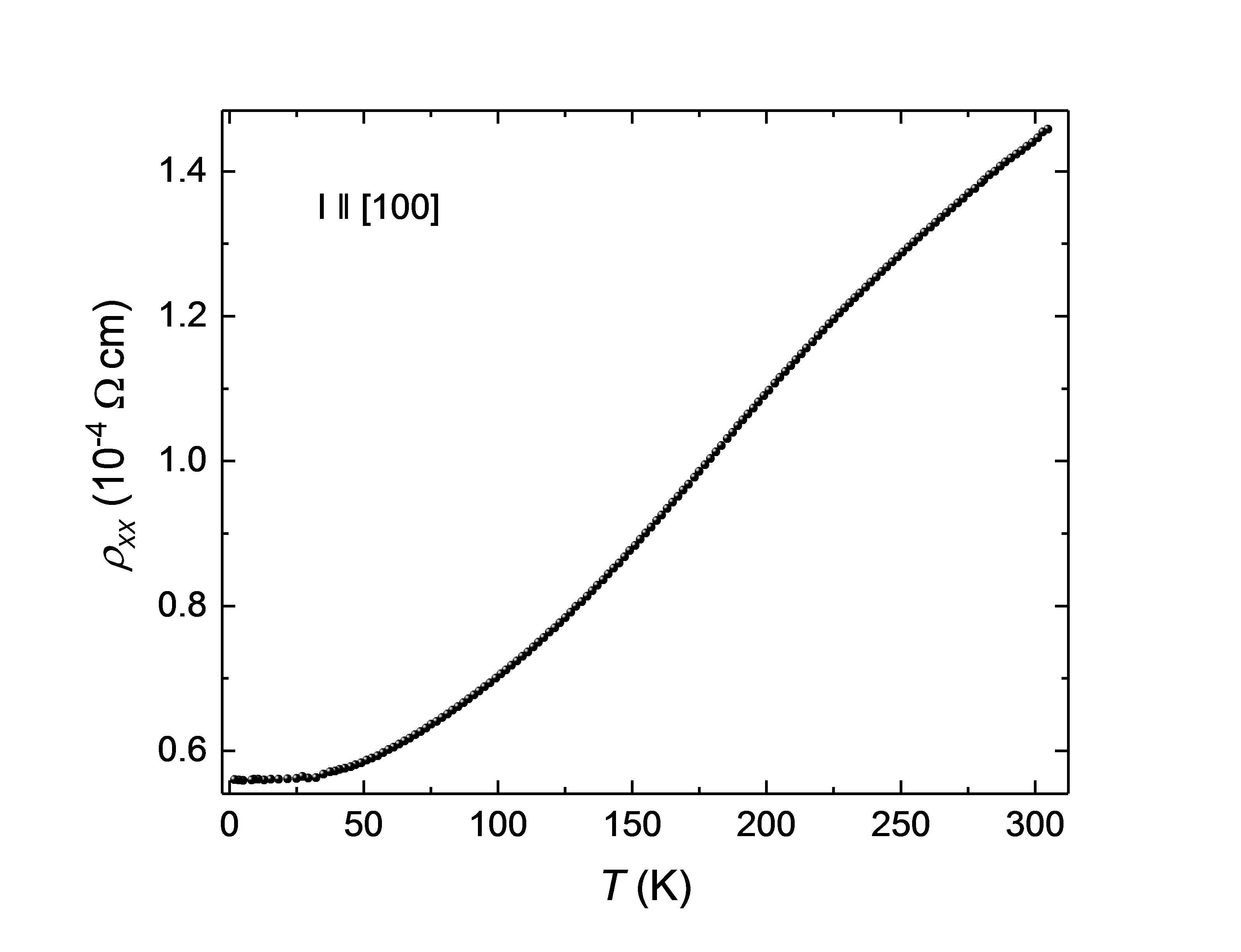}
\caption{\label{FigS3} \textbf{Temperature dependent resistivity.} Longitudinal resistivity as a function of temperature for a Co$_2$MnGa single crystal with current along the [100] direction.}
\end{figure*}

\begin{figure*}[h]
\centering
\includegraphics[width=16cm,trim={1in 7in 1in 1in},clip]{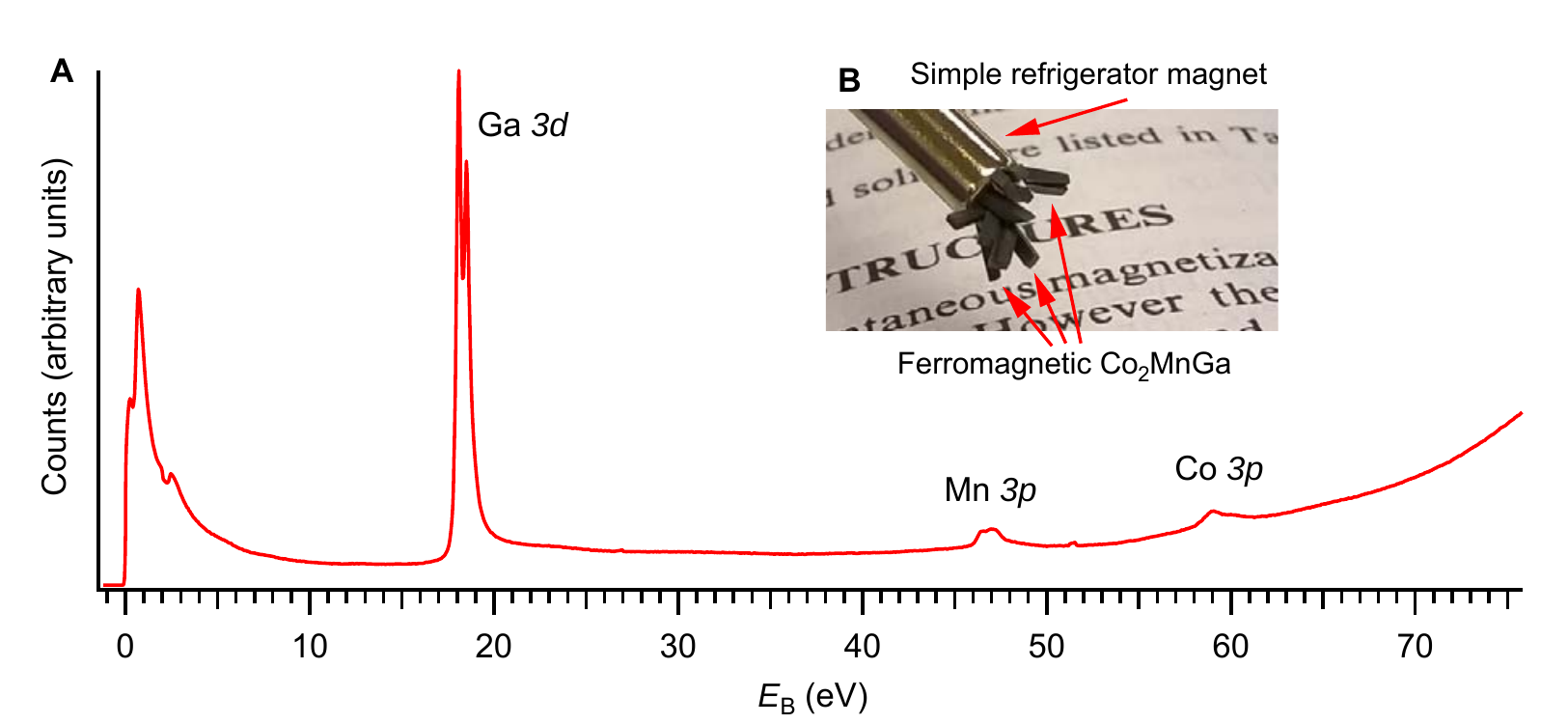}
\caption{\label{FigCore} \textbf{Core level spectrum of Co$_2$MnGa.} \textbf{A}, An XPS spectrum of our Co$_2$MnGa samples clearly shows Ga $3d$, Mn $3p$ and Co $3p$ peaks, without significant irrelevant core level peaks, suggesting that our samples are of high quality. \textbf{B}, The single crystal Co$_2$MnGa samples are readily picked up by an ordinary refrigerator magnet at room temperature.}
\end{figure*}

\begin{figure*}[h]
\centering
\includegraphics[width=16cm,trim={1in 4.5in 1in 1in},clip]{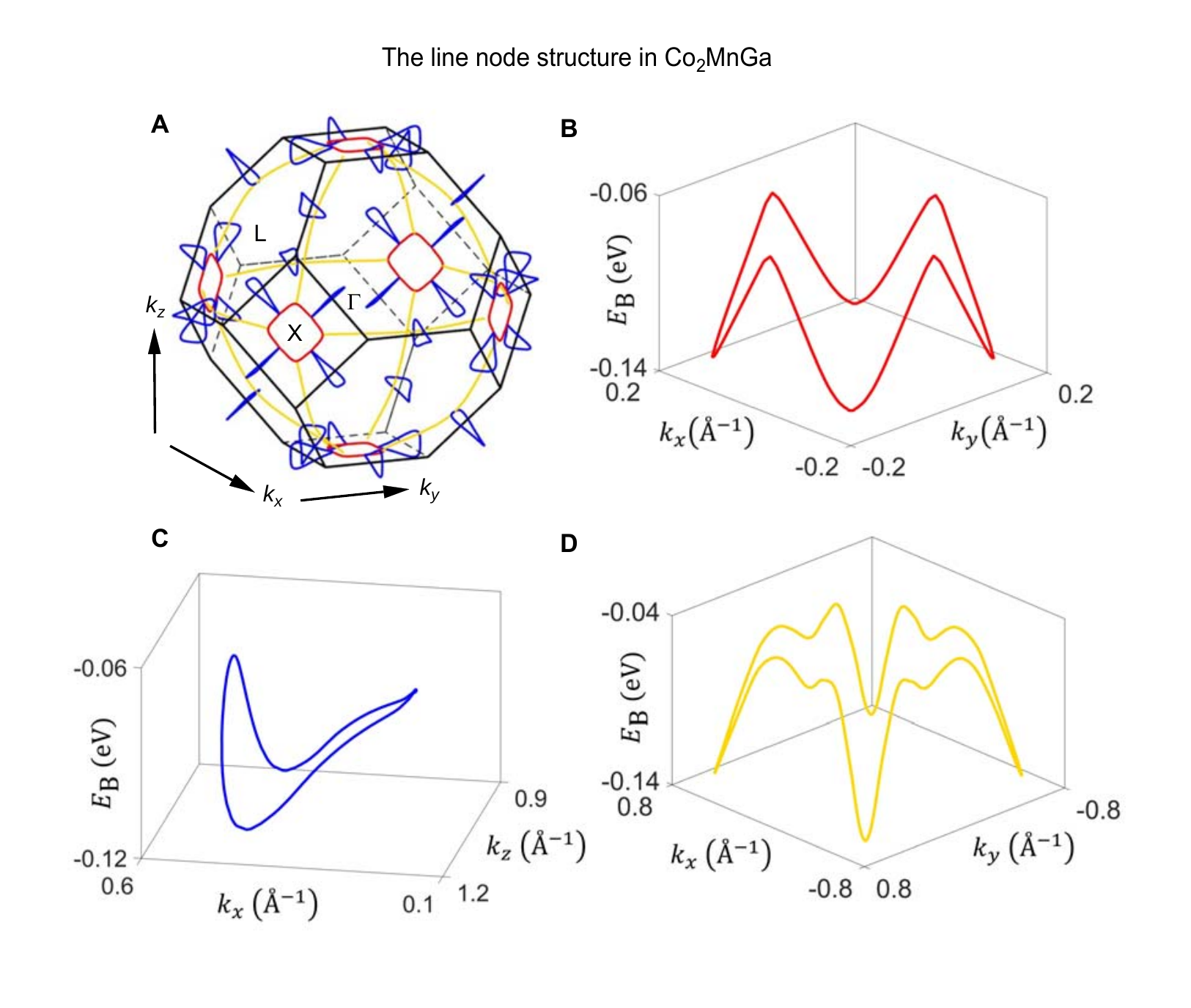}
\caption{\label{FigCalc} \textbf{Dispersions of the line nodes in Co$_2$MnGa.} \textbf{A}, Plot of the line nodes in the bulk Brillouin zone in Co$_2$MnGa, same as main text Fig. 1e, repeated here again for clarity. Although they are locked on a mirror plane, the line nodes have a dispersion in energy, shown in \textbf{B}-\textbf{C} for the red, blue and yellow line nodes, respectively. All line nodes are generated from these three by applying symmetry operations of the crystal lattice, generating the complicated line node network shown in \textbf{A}.}
\end{figure*}

\begin{figure*}[h]
\centering
\includegraphics[width=16cm,trim={0.8in 0.8in 0.8in 0.8in},clip]{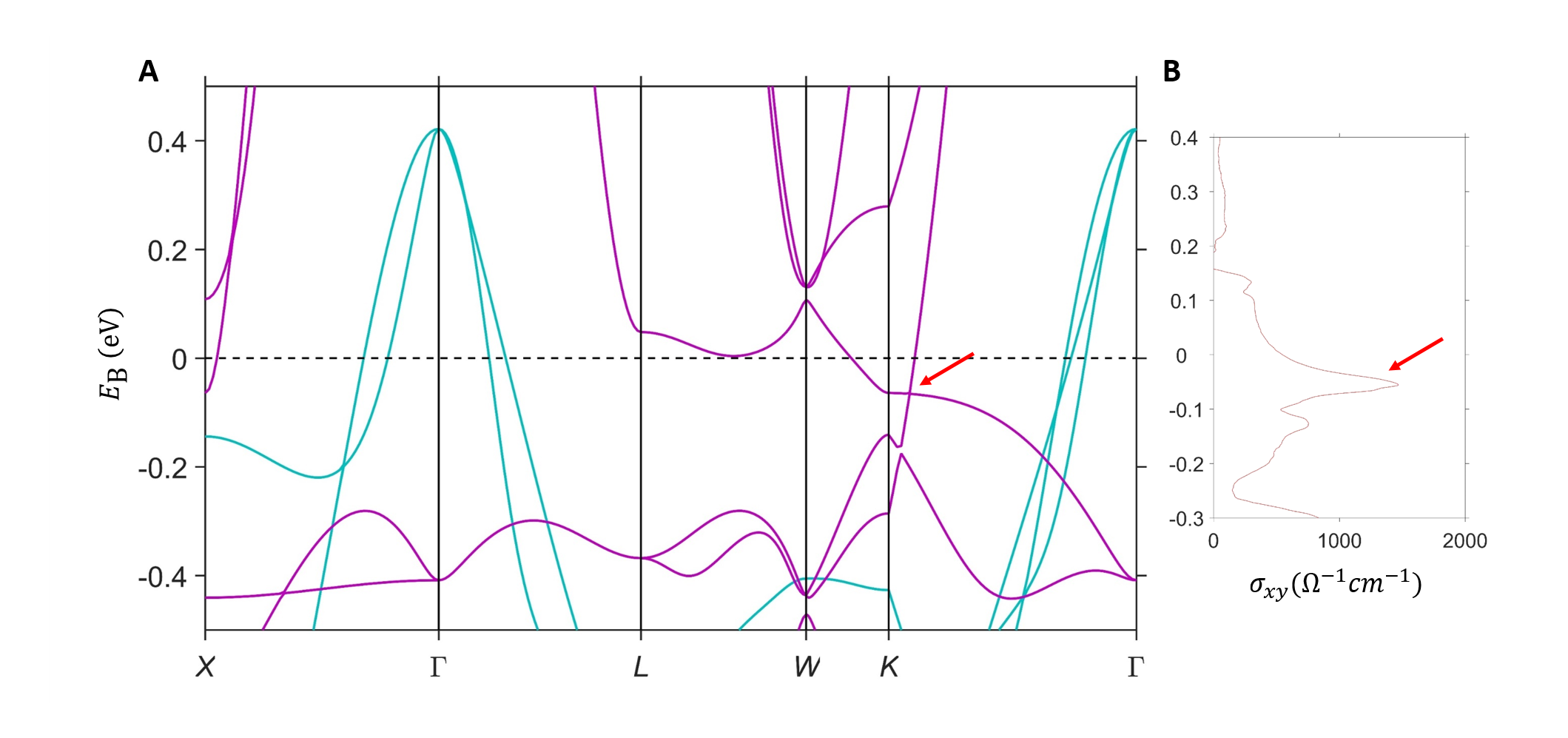}
\caption{\label{FigBerry} \textbf{Berry curvature distribution of Co$_2$MnGa} \textbf{A}, The same panel as main text Fig. 1B, repeated here for convenience. \textbf{B}, The integral of the Berry curvature up to a particular binding energy. We observe a large peak at the energy of the line nodes (red arrow), suggesting that the line nodes dominate the Berry curvature in the system and are responsible for the anomalous Hall response.}
\end{figure*}

\begin{figure*}[h]
\centering
\includegraphics[width=14cm,trim={0in 0in 0in 0in},clip]{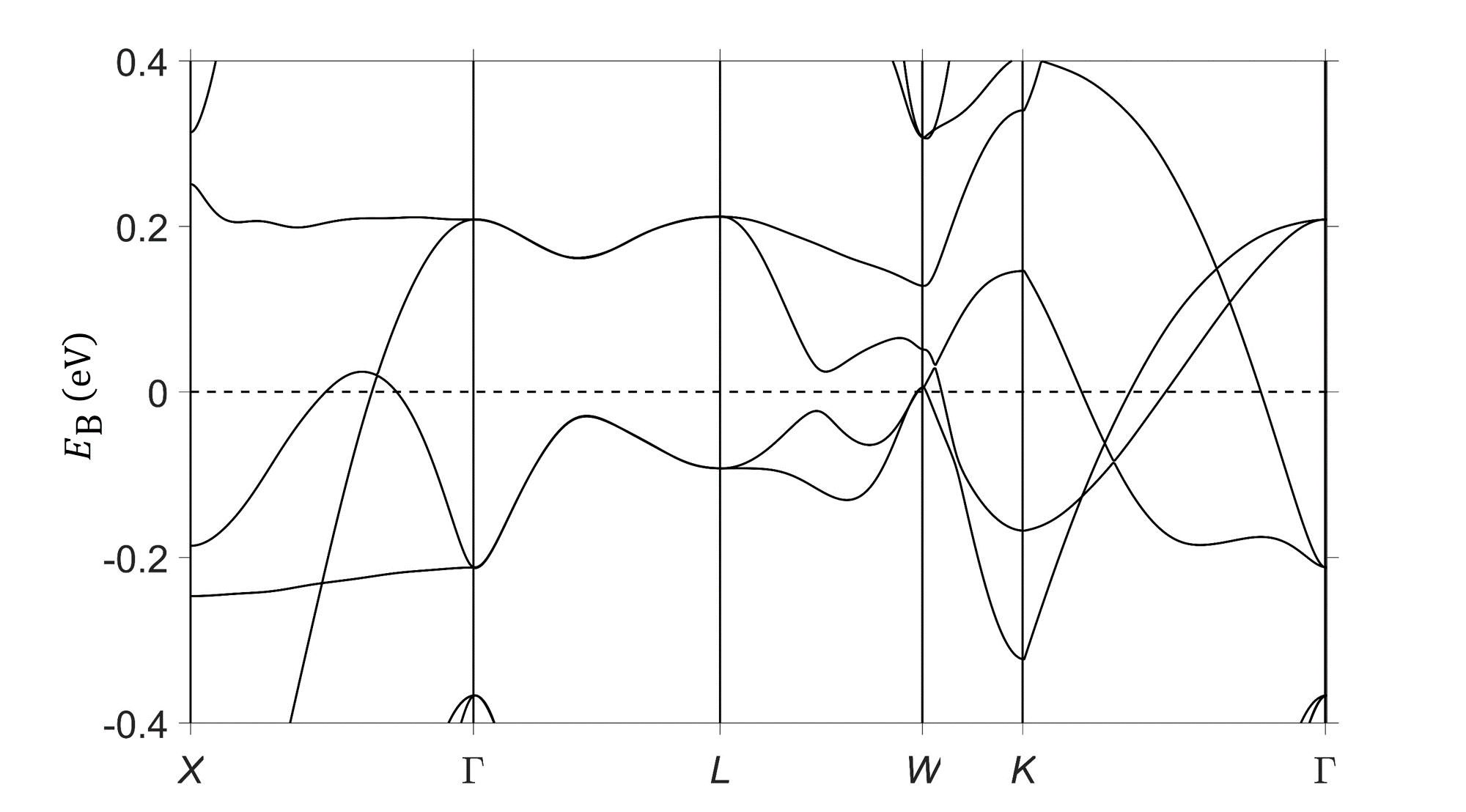}
\caption{\label{FigNonMag} \textbf{Non-magnetic calculation of the band structure of Co$_2$MnGa.} Without magnetism, the band structure changes completely and the line nodes disappear, showing that magnetism is essential for the topological phase of the system. Note, however, that new, unrelated line nodes may arise in the non-magnetic band structure.}
\end{figure*}

\begin{figure*}[h]
\centering
\includegraphics[width=16cm,trim={1.4in 3.7in 1.4in 1.4in},clip]{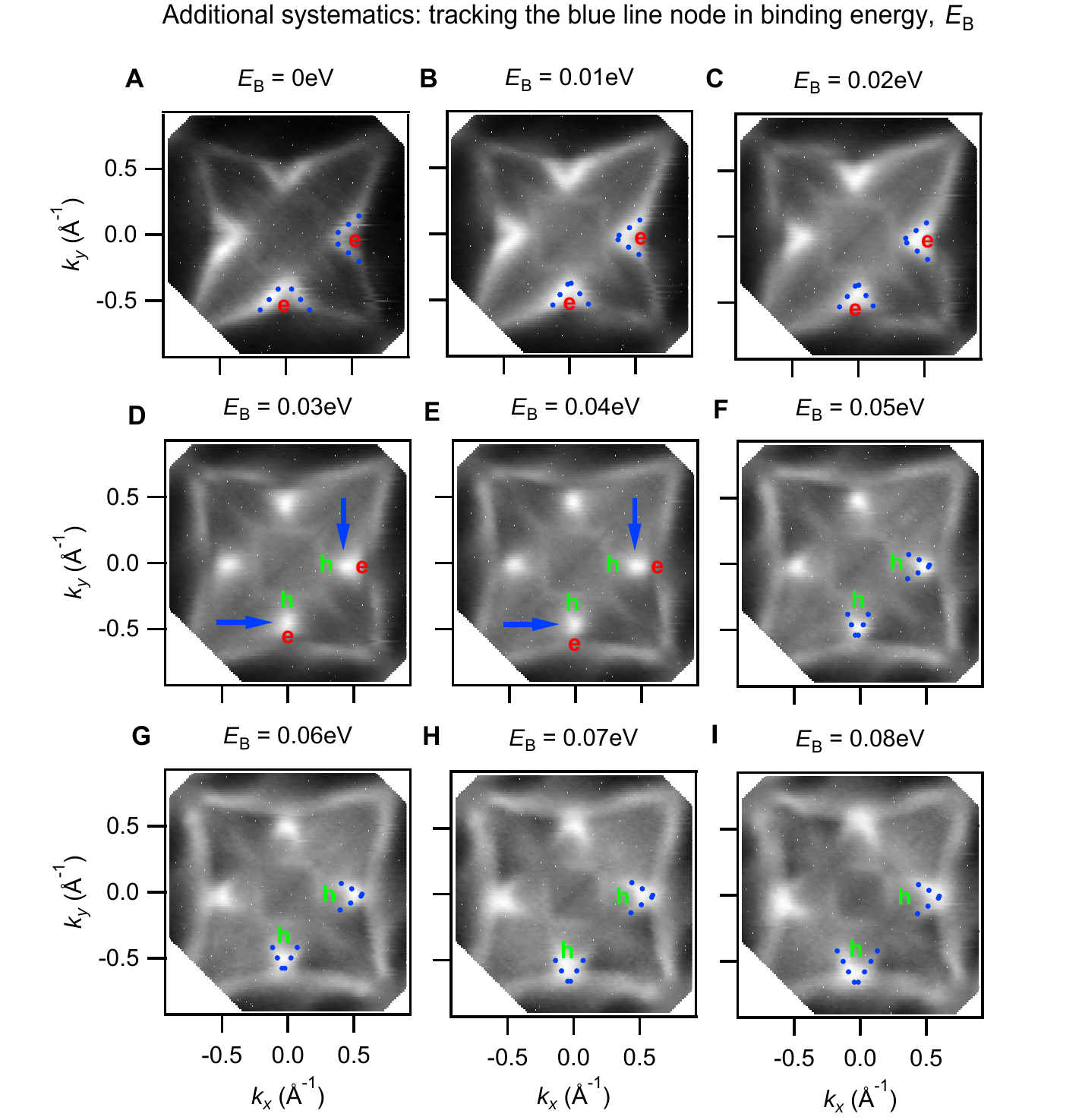}
\caption{\label{FigBlueFS} \textbf{Constant energy surfaces of Co$_2$MnGa.} \textbf{A}-\textbf{I}, We expand on the dataset shown in main text Fig. 3A-D by plotting constant energy surfaces at additional binding energies. This allows us to more smoothly see the transition from electron to hole pocket as we sweep through the line node in energy. This behavior of the band structure demonstrates a line node in experiment, essentially without relying on calculation.}
\end{figure*}

\begin{figure*}[h]
\centering
\includegraphics[width=16cm,trim={1.2in 3.7in 1.2in 1.2in},clip]{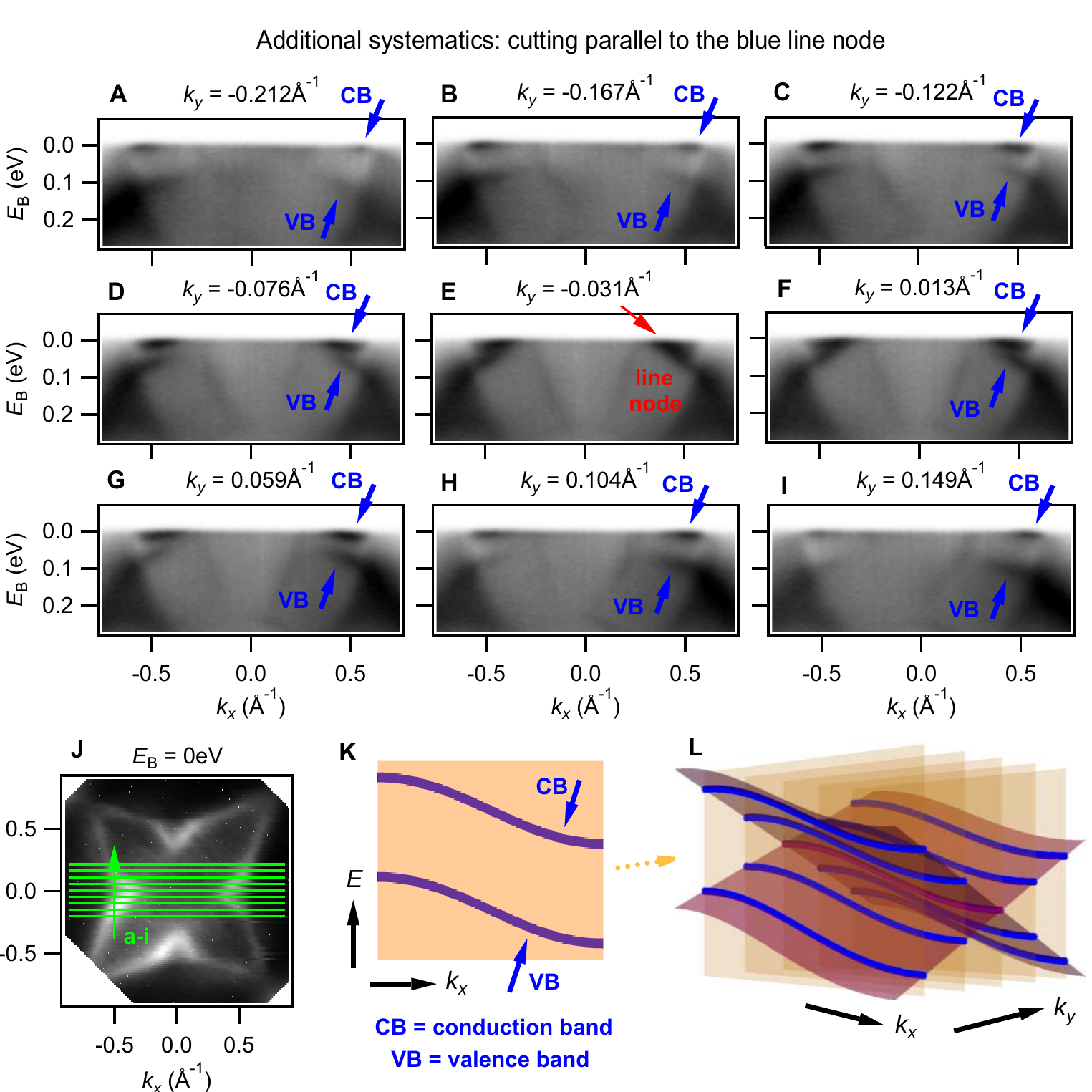}
\caption{\label{FigBluePar} \textbf{A different perspective on the blue line node.} \textbf{A}-\textbf{I}, $E_\textrm{B}-k_x$ cuts sweeping through the line node in $k_y$, as marked on the constant energy surface in \textbf{J}. We clearly observe that the valence and conduction bands approach each other and meet on a finite line in momentum space, illustrated by the red arrow in \textbf{E}. This again demonstrates a line node in Co$_2$MnGa in ARPES. Note that in our data the valence and conduction bands in fact meet slightly away from $k_y = 0 \textrm{\AA}^{-1}$, probably due to a small error in the calibration of the momentum axes or a slight sample misalignment. \textbf{K}, Schematic showing a generic spectrum cutting parallel to the line node, as well as its evolution in $k_y$, \textbf{L}.}
\end{figure*}

\begin{figure*}[h]
\centering
\includegraphics[width=16cm,trim={1.2in 1.2in 1.2in 1.2in},clip]{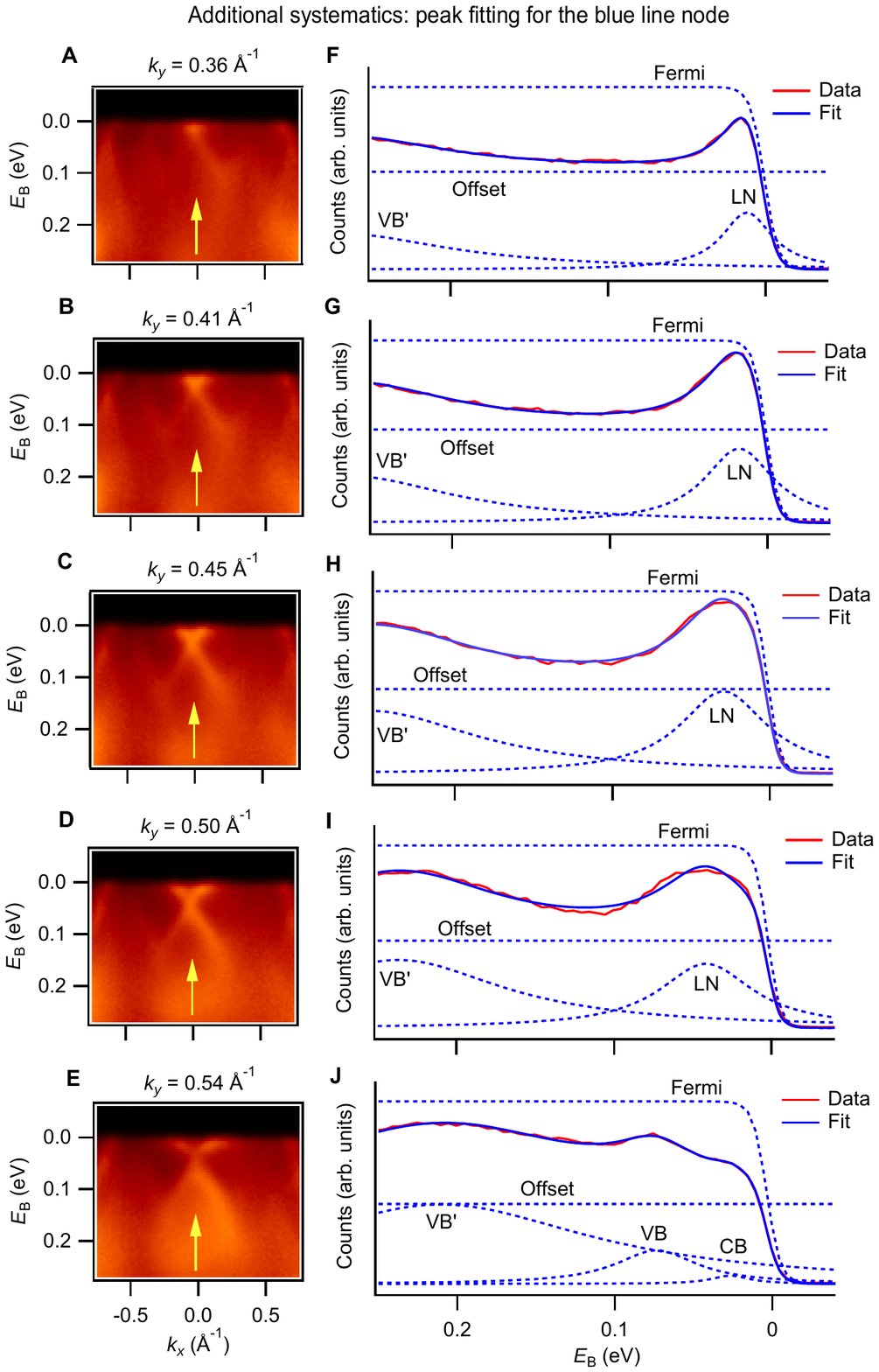}
\end{figure*}

\clearpage
\begin{figure*}[h]
\caption{\label{FigBlueFit} \textbf{Demonstrating a band crossing through fitting.} \textbf{A}-\textbf{E}, $E_\textrm{B}-k_x$ cuts sweeping perpendicular to the line node, the same as in main text Fig. 2E-I, shown here again for clarity. To distinguish between a small gap and a crossing we fit the relevant energy distribution curve (EDC) to a Lorentzian peak. The excellent match for \textbf{F}-\textbf{H} shows a line node. For \textbf{I} the fit worsens possibly because we are at the edge of the line node projection. Lastly, for \textbf{J} we find twin peaks emerge, again each well-described by a Lorentzian, corresponding to a gap for cuts beyond the line node. The behavior provides additional evidence for a line node in Co$_2$MnGa.}
\end{figure*}

\clearpage
\begin{figure*}[h]
\centering
\includegraphics[width=16cm,trim={1.2in 6.2in 1.2in 1.2in},clip]{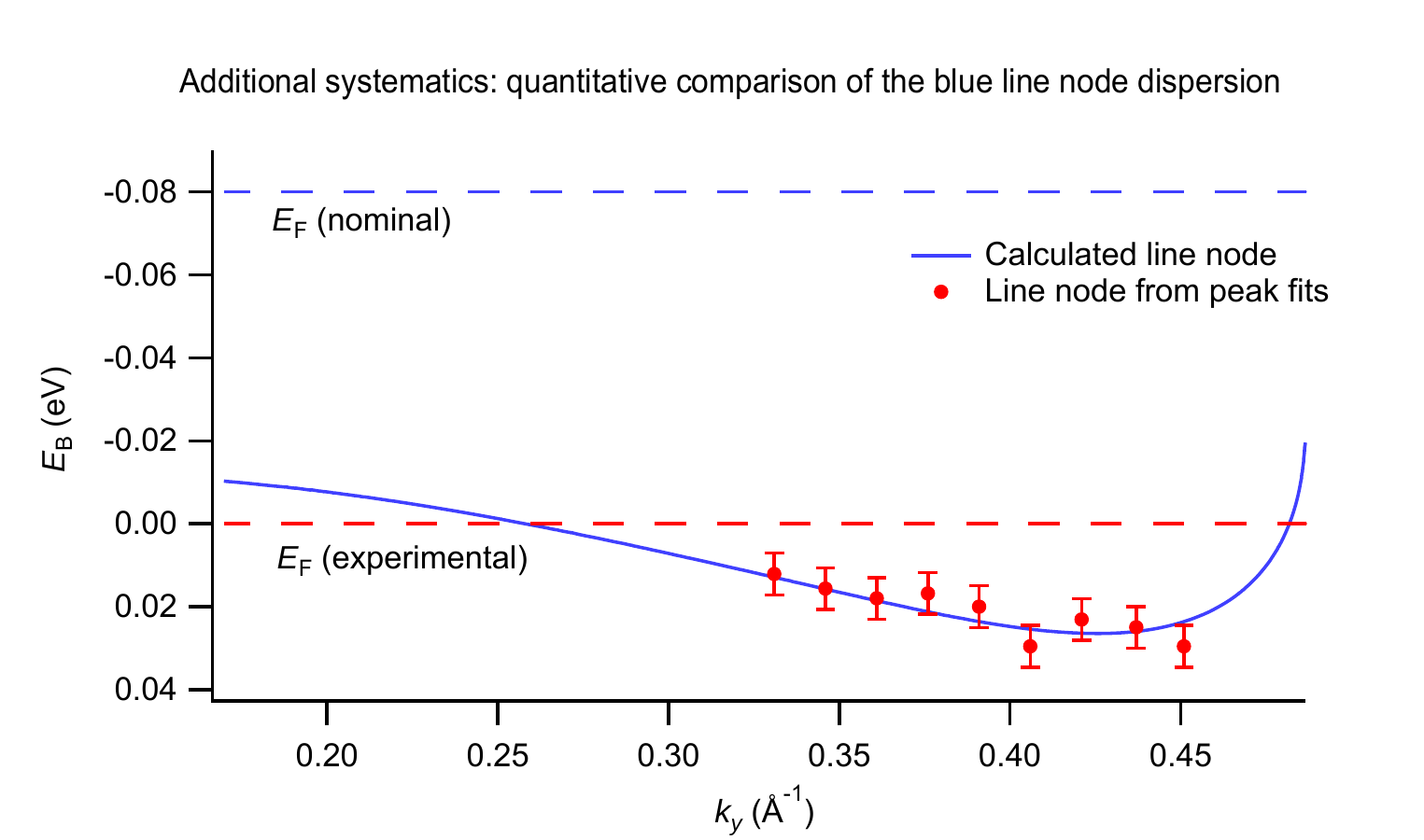}
\caption{\label{FigBlueTrack} \textbf{Comparing Lorentzian peaks to the calculated line node.} We track the line node from the Lorentzian peak fits in Fig. \ref{FigBlueFit} and superimpose the calculated line node dispersion, with a $0.08$ eV shift to account for hole doping in the sample, as discussed in the main text. We find a reasonable match between Lorentzian fits of the line node and \textit{ab initio} calculation.}
\end{figure*}

\begin{figure*}[h]
\centering
\includegraphics[width=16cm,trim={1.2in 2.2in 1.2in 1.2in},clip]{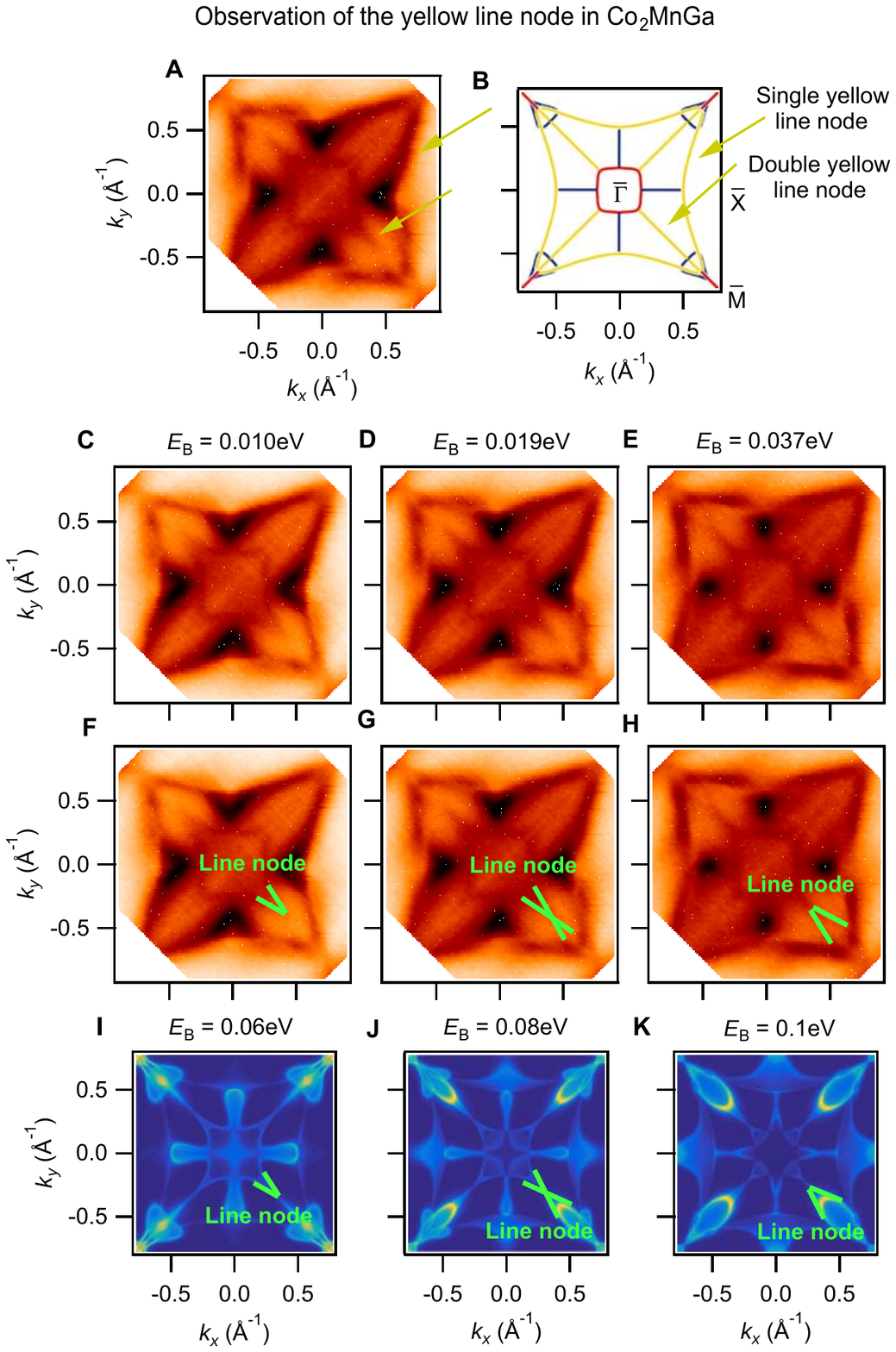}
\end{figure*}

\clearpage
\begin{figure*}[h]
\caption{\label{FigYellow1} \textbf{Characteristic evolution of the constant energy surface for the double yellow line node.} By comparing a measured constant energy surface, \textbf{A}, with the (001) projection of the line nodes from calculation, \textbf{B}, we find signatures of the yellow line nodes in Co$_2$MnGa. We note different yellow line nodes project differently on the (001) surface. The four yellow line nodes emmanating out from $\bar{\Gamma}$ project ``standing up'' so that the band crossings project in pairs into the surface Brillouin zone, giving four so-called ``double'' yellow line nodes. The remaining yellow line node forms a large ring around the entire surface Brillouin zone, projecting simply ``face up'', giving a ``single'' yellow line node. \textbf{C}-\textbf{E}, constant energy surfaces where we directly observe the characteristic $<$ to $>$ transition, as marked in \textbf{F}-\textbf{H}. \textbf{I}-\textbf{K}, We observe the same evolution in calculated constant energy surfaces, consistent with our experimental results, demonstrating the double yellow line node. We emphasize that we cannot determine from our data alone whether the line node is single or double, but we can demonstrate that there is some type of line node crossing.}
\end{figure*}

\clearpage
\begin{figure*}[h]
\centering
\includegraphics[width=16cm,trim={1.2in 3.1in 1.2in 1.3in},clip]{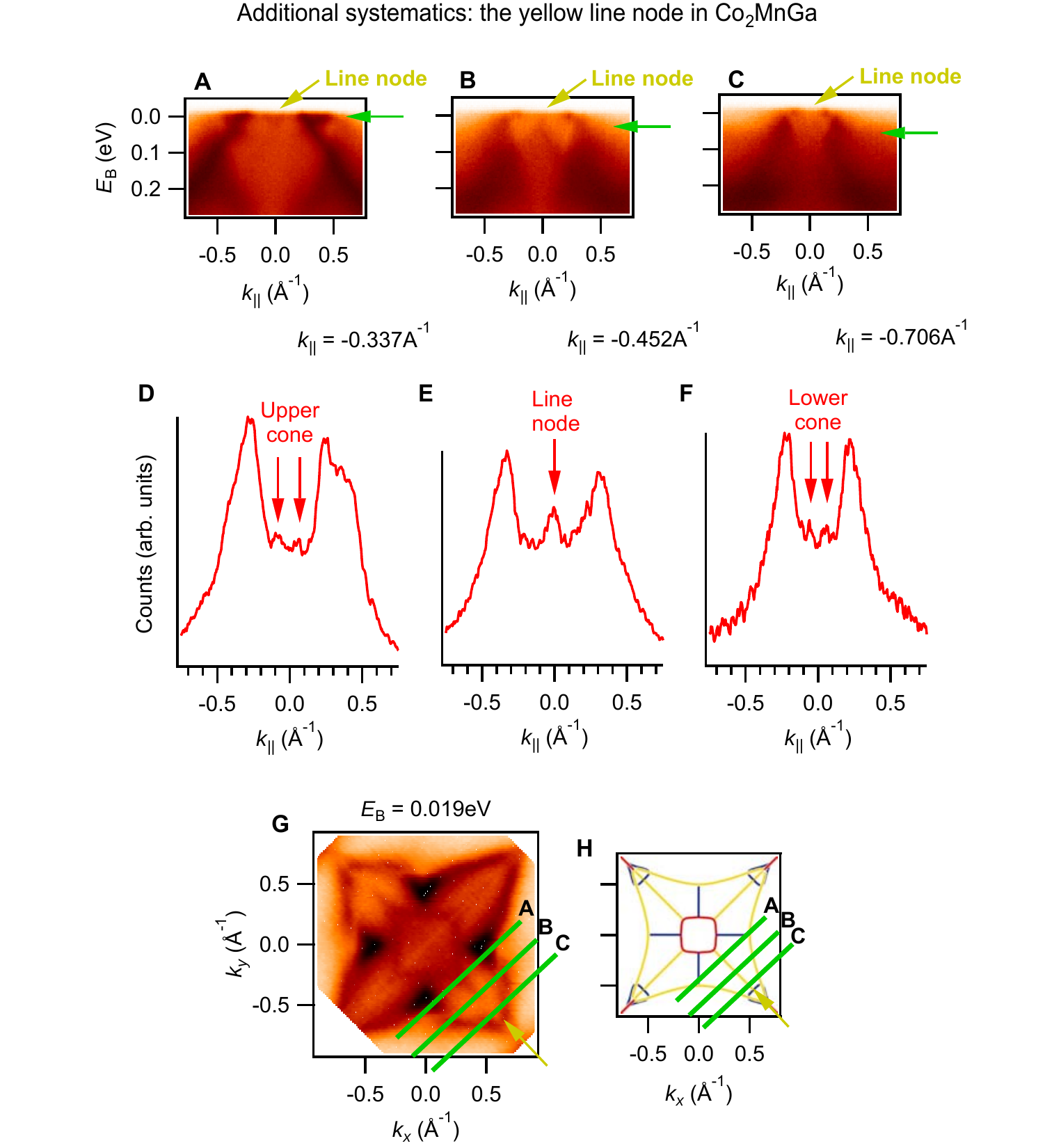}
\caption{\label{FigYellow2} \textbf{Pinpointing the yellow line node in momentum distribution curves (MDCs).} \textbf{A}-\textbf{C}, $E_\textrm{B}-k_{||}$ cuts sweeping perpendicular to the line node. In \textbf{A} we observe the upper cone associated with the double yellow line node; in \textbf{B} we see a line node crossing and the lower cone; in \textbf{C} we find a weak signature remaining from the lower cone. \textbf{D}-\textbf{E}, MDCs taken from the $E_\textrm{B}-k_{||}$ cuts, as indicated by the green arrows in \textbf{A}-\textbf{C}. The weak peaks associated with the double yellow line node are marked by the red arrows. \textbf{G}, \textbf{H}, The locations of the cuts in \textbf{A}-\textbf{C}, as marked. Our MDC analysis provides additional evidence for the yellow line node in Co$_2$MnGa.}
\end{figure*}

\begin{figure*}[h]
\centering
\includegraphics[width=16cm,trim={1.1in 1.5in 1.1in 1.2in},clip]{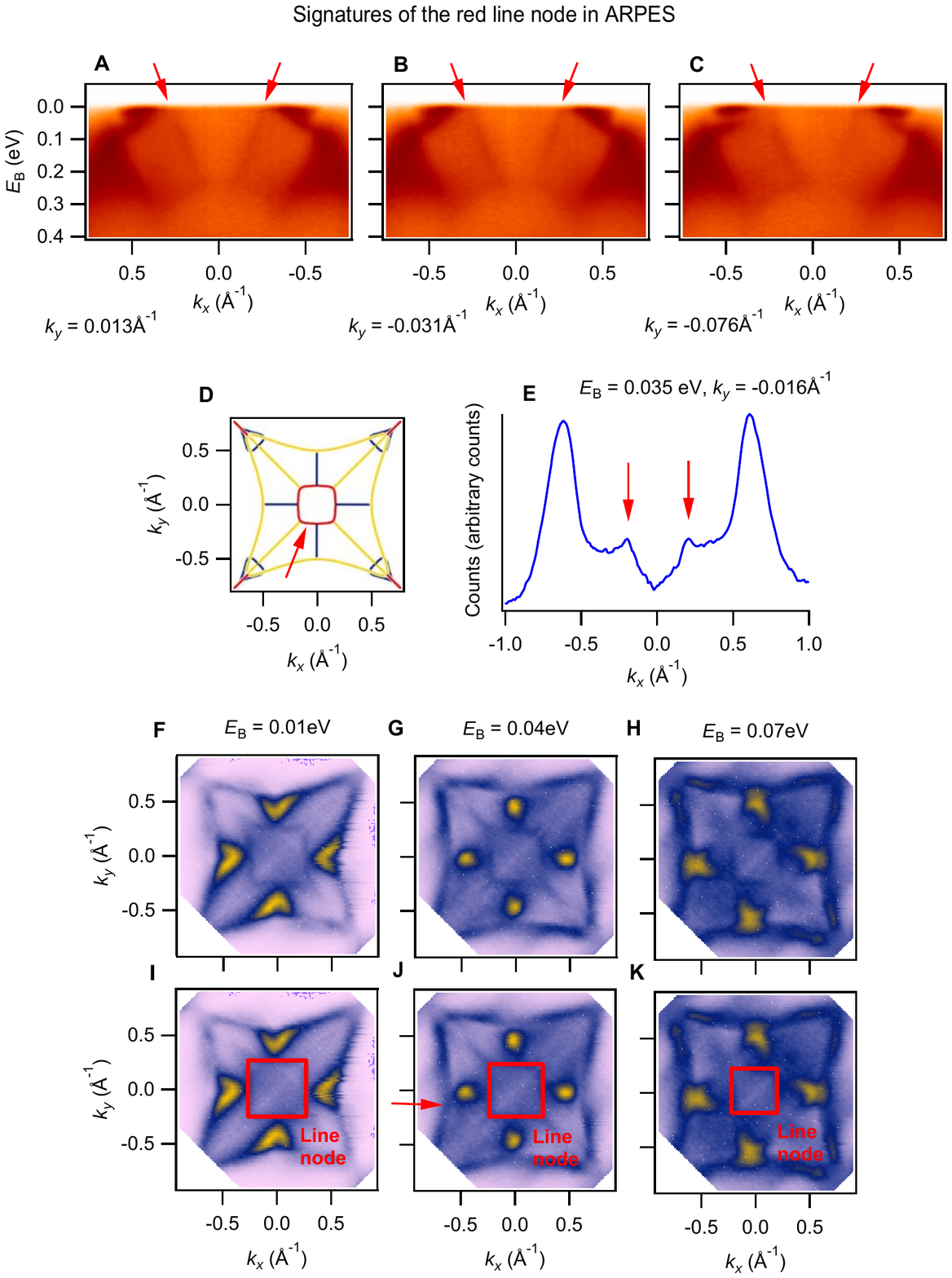}
\end{figure*}

\clearpage
\begin{figure*}[h]
\caption{\label{FigRed} \textbf{Single branch of the red line node.} \textbf{A}-\textbf{C}, $E_\textrm{B}-k_x$ cuts passing through the red line node near $\bar{\Gamma}$. We see two bands dispersing away from $\bar{\Gamma}$ as we approach $E_\textrm{F}$. \textbf{D}, The red line node in calculation. \textbf{E}, The two bands of interest, pinpointed on a momentum distribution curve (MDC). \textbf{F}-\textbf{H}, Constant energy surfaces, emphasizing a square feature around $\bar{\Gamma}$ in agreement with the red line node as pinpointed in \textbf{D}, with the square feature marked in \textbf{I}-\textbf{K} and the location of the MDC shown by the red arrow in \textbf{J}. The square feature provides a clear signature of the red line node, although it appears that only one branch of the crossing is visible, so we can not directly observe the line node cone.}
\end{figure*}

\clearpage
\begin{figure*}[h]
\centering
\includegraphics[width=16cm,trim={1.1in 5.5in 1.1in 1.2in},clip]{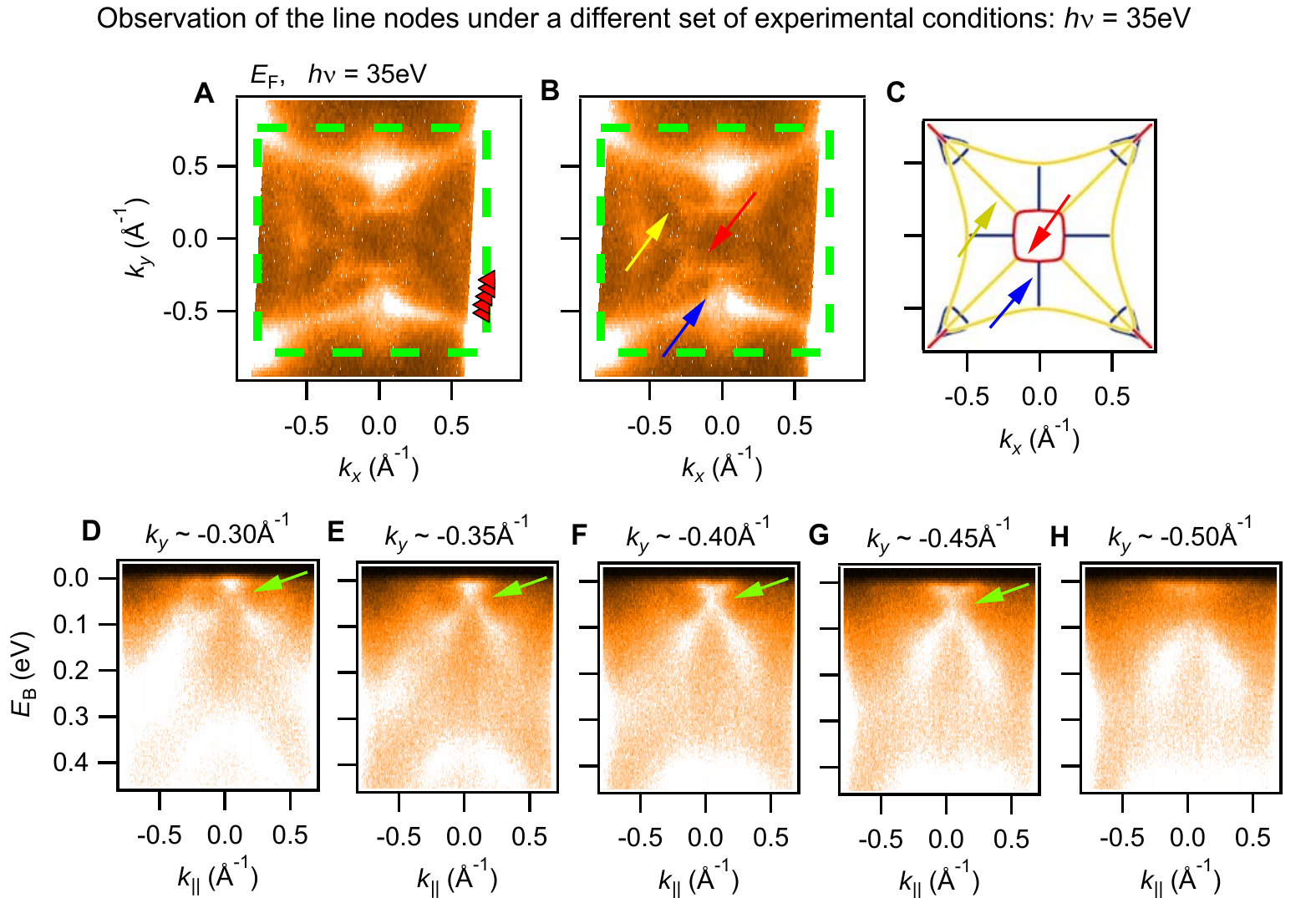}
\caption{\label{Fig35eV} \textbf{The blue line node at a different incident photon energy.} \textbf{A}, Fermi surface at $h\nu = 35$ eV and \textbf{B}, the same Fermi surface with the line nodes marked by the arrows, corresponding to the arrows in \textbf{C}. \textbf{D}-\textbf{H}, We directly observe the blue line node, consistent with calculation and our data at $h\nu = 50$ eV. We find a clear gap as we cut past the end of the line node projection, \textbf{H}, again consistent with our earlier discussion. These results provide yet another independent check of our observation of a line node in Co$_2$MnGa.}
\end{figure*}

\clearpage
\begin{figure*}[h]
\centering
\includegraphics[width=16cm,trim={1.1in 5.5in 1.1in 1.2in},clip]{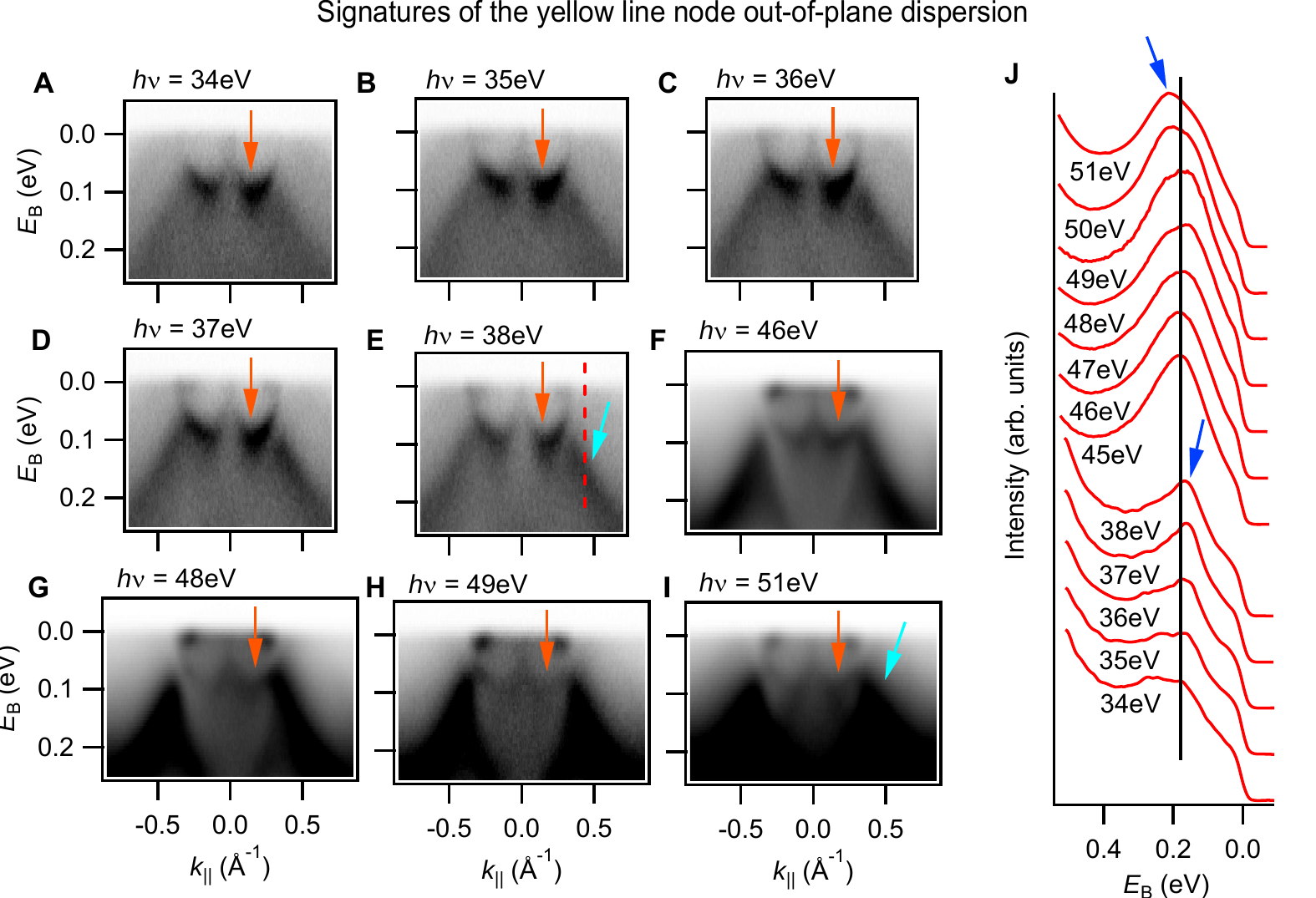}
\caption{\label{DH1} \textbf{Full photon energy dependence of the drumhead and yellow line node.} \textbf{A-I}, $E_\textrm{B}-k_{||}$ cuts analagous to main text Fig. 4A-C, but at more photon energies. \textbf{J}, Stack of energy distribution curves (EDCs) as a function of photon energy, analogous to main text Fig. 4G, but instead of cutting through the drumhead surface state, the EDC cuts through the yellow line node, at $k_{||} = 0.45\textrm{\AA}^{-1}$ (dotted red line in \textbf{E}). We clearly observe the drumhead surface state in all cuts (orange arrows). Recall that the drumhead showed no photon energy dependence, suggesting that it is a surface state. Here, by contrast, we see a clear photon energy dependence (blue arrows in \textbf{J}), associated with the yellow line node cones (cyan arrows in \textbf{E}, \textbf{I}). The photon energy dependence suggests a $k_z$ dispersion for the yellow line node cone, as expected.}
\end{figure*}

\clearpage
\begin{figure*}[h]
\centering
\includegraphics[width=16cm,trim={1in 6.8in 1in 1.2in},clip]{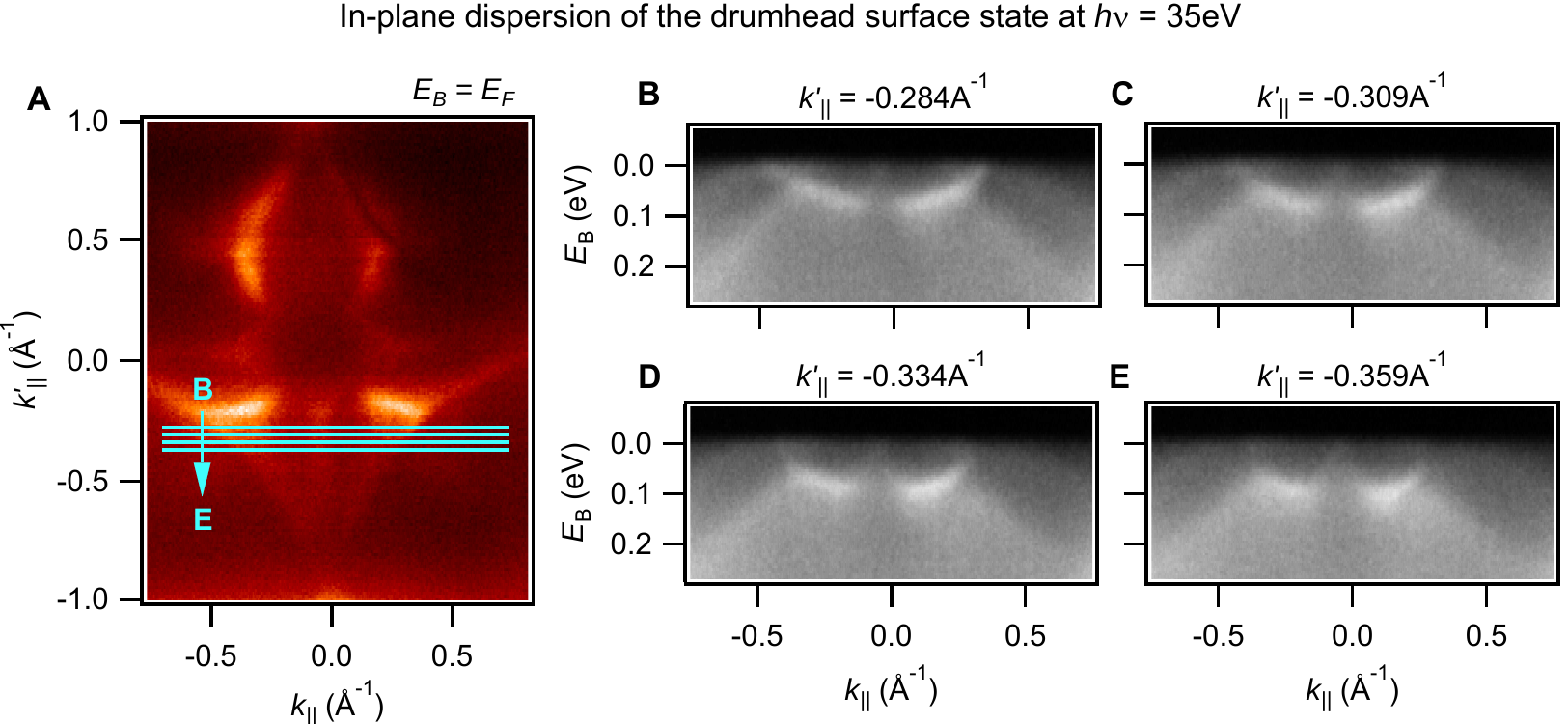}
\caption{\label{DH2} \textbf{In-plane dispersion of the drumhead.} \textbf{A}, Fermi surface at $h\nu = 35$ eV and \textbf{B-E}, $E_\textrm{B}-k_{||}$ cuts showing the drumhead surface state, with locations as marked in \textbf{A} (cyan lines). We observe a weak dispersion of the surface state doward in energy as we move away from $\bar{\Gamma}$. So we a observe an in-plane dispersion of the drumhead, but not an out-of-plane dispersion (main text Fig. 4), demonstrating a surface state.}
\end{figure*}

\clearpage
\begin{figure*}[h]
\centering
\includegraphics[width=16cm,trim={1.1in 2.2in 1.1in 1.05in},clip]{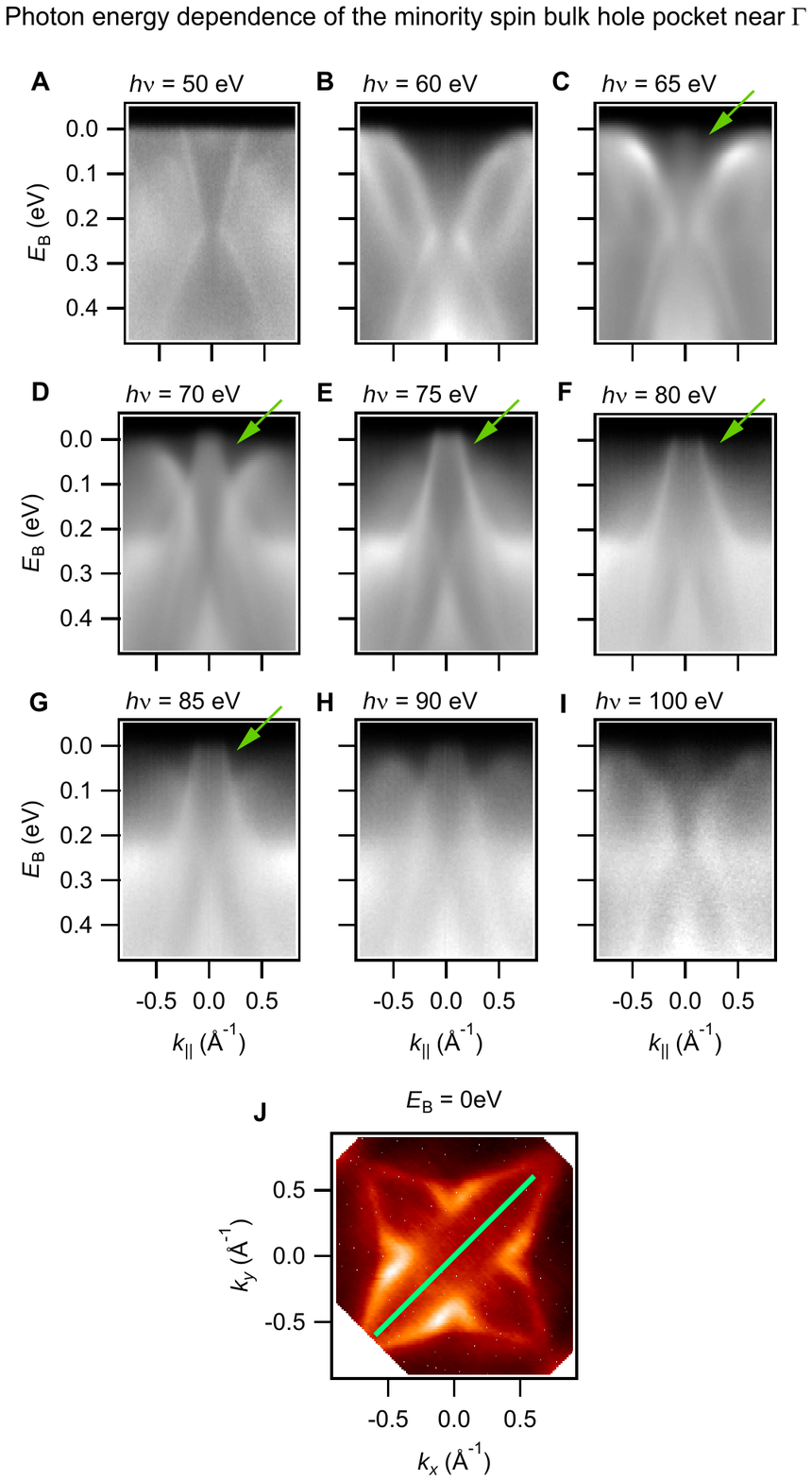}
\caption{\label{spinmin} \textbf{The irrelevant minority spin pocket.} \textbf{A-I}, $E_\textrm{B}-k_{||}$ cuts through $\bar{\Gamma}$ at different photon energies, with the location of the cut shown in \textbf{J} (green line). At  $h \nu > 65$ eV a large, clear hole pocket appears (green arrows), consistent with the minority spin pocket observed in calculation. This result suggests that the minority spin pocket was suppressed in our measurements at $h \nu = 50$ eV because we were cutting at $k_z \sim \pi$, far from $k_z \sim 0$.}
\end{figure*}

\end{document}